\documentclass[lettersize,journal]{IEEEtran}
\usepackage{amsmath,amsfonts}
\usepackage{amsthm}
\usepackage{algorithmic}
\usepackage{algorithm}
\usepackage{array}
\usepackage{amssymb}
\usepackage{bm}

\usepackage[caption=false,font=footnotesize,labelfont=rm,textfont=rm]{subfig}
\usepackage{titlesec}
\usepackage{color}
\usepackage{xcolor}
\usepackage{graphicx}
\usepackage{textcomp}
\usepackage{stfloats}
\usepackage{url}
\usepackage{verbatim}
\usepackage{graphicx}
\usepackage{epstopdf}
\usepackage{cite}
\newtheorem{proposition}{\textbf{Proposition}}
\begin{document}

\title{
	\huge{Low-Complexity Channel Estimation for Internet of Vehicles AFDM Communications
	With Sparse Bayesian Learning}
}

\author{Xiangxiang Li, Haiyan Wang, \emph{Member, IEEE}, Yao Ge, \emph{Member, IEEE},  Xiaohong Shen, \emph{Member, IEEE},  \\ Miaowen Wen,  \emph{Senior Member, IEEE}, Shun Zhang, \emph{Senior Member, IEEE}, and Yong Liang Guan,  \\ \emph{Senior Member, IEEE}
	
	\thanks{This work was supported by the National Natural Science Foundations of China under Grant No.62031021 and No.62271404.
	The work of Yao Ge is supported by A*STAR under the RIE2025 Industry Alignment Fund - Industry Collaboration Projects (IAF-ICP) Funding Initiative (Award: I2501E0045), as well as cash and in-kind contribution from the industry partner(s). (\emph{Corresponding authors: Haiyan Wang; Yao Ge}.)}
	\thanks{Xiangxiang Li, Haiyan Wang and Xiaohong Shen are with the School of Marine Science and Technology and the	Key Laboratory of Ocean Acoustics and Sensing, Ministry of Industry and Information Technology, Northwestern Polytechnical University, Xi’an, Shaanxi 710072, China. Haiyan Wang is also with the School of Electronic Information and Artifcial Intelligence, Shaanxi University of Science and Technology,	Xi’an, Shaanxi 710021, China (e-mail: lixx@mail.nwpu.edu.cn; hywang@nwpu.edu.cn; xhshen@nwpu.edu.cn).}
	\thanks{Yao Ge is with the AUMOVIO-NTU Corporate Lab, Nanyang Technological University, Singapore 639798 (e-mail: yao.ge@ntu.edu.sg).}
	\thanks{Miaowen Wen is with the School of Electronic and Information Engineering, South China University of Technology, Guangzhou 510641, China (e-mail: eemwwen@scut.edu.cn).}
	\thanks{Shun Zhang is with the State Key Laboratory
	of Integrated Services Networks, Xidian University, Xi'an 710071, China (e-mail: zhangshunsdu@xidian.edu.cn).}
	\thanks{Yong Liang Guan is with the School of Electrical and Electronics Engineering, Nanyang Technological University, Singapore 639798 (e-mail: eylguan@ntu.edu.sg).}
}

\maketitle
\setlength{\textfloatsep}{4.5pt}

\begin{abstract}
Affine frequency division multiplexing (AFDM) has been considered as a promising waveform to enable high-reliable connectivity in the internet of vehicles. However, accurate channel estimation is critical and challenging to achieve the expected performance of the AFDM systems in doubly-dispersive channels.
In this paper, we propose a sparse Bayesian learning (SBL) framework for AFDM systems and develop a dynamic grid update strategy with two off-grid channel estimation methods, i.e., grid-refinement SBL (GR-SBL) and grid-evolution SBL (GE-SBL) estimators.
Specifically, the GR-SBL employs a localized grid refinement method and 
dynamically updates grid for a high-precision estimation. 
The GE-SBL estimator approximates the off-grid components via first-order linear approximation and enables gradual grid evolution for estimation accuracy enhancement.
Furthermore, we develop a distributed computing scheme to decompose the large-dimensional channel estimation model into multiple manageable small-dimensional sub-models for complexity reduction of GR-SBL and GE-SBL, denoted as distributed GR-SBL (D-GR-SBL) and distributed GE-SBL (D-GE-SBL) estimators, which also support parallel processing to reduce the computational latency.
Finally, simulation results demonstrate that the proposed channel estimators outperform existing competitive schemes. 
The GR-SBL estimator achieves high-precision estimation with fine step sizes at the cost of high complexity, while the GE-SBL estimator provides a better trade-off between performance and complexity. 
The proposed D-GR-SBL and D-GE-SBL estimators effectively reduce complexity and maintain comparable performance to GR-SBL and GE-SBL estimators, respectively.
\end{abstract}


\section{Introduction}
The beyond-5G and 6G wireless communication networks are considered as a promising enabler of the internet of vehicles. As 
a key application of intelligent transportation systems, the internet of vehicles are excepted to support the low-latency and high-reliable connectivity in the diverse high-mobility scenarios.
%The rapid evolution of wireless communication technologies has driven the widespread adoption of the internet of things (IoT), which aims to enable seamless connectivity for everything at any time and place. As a key application of IoT in intelligent transportation systems, vehicle-to-everything (V2X) is excepted to deliver the low-latency and ultra-reliable connectivity in diverse high-mobility scenarios. 
%The rapid evolution of wireless communication technologies has driven the widespread adoption of the internet of things (IoT), which is excepted to support the ultra-reliable connectivity in diverse high-mobility scenarios, such as vehicle network, low-earth-orbit satellites (LEOS), unmanned aerial vehicles (UAV) and so on. 
However, the relative motion between transmitters and receivers induces severe Doppler spread that inevitably disrupts the sub-carrier orthogonality in conventional orthogonal frequency division multiplexing (OFDM) \cite{b1}. This degradation generates inter-carrier interference (ICI), leading to severe performance loss. Therefore, the high-reliable internet of vehicles communication over doubly-dispersive channels has become important and challenging for next-generation wireless communication systems.

%aims to connect everything
%
%
%The internet of thing (IoT) are expected to support ultra-reliable connectivity in high-mobility scenarios, such as 
%
%high speed-trains, low-Earth-orbit satellites
%
%
%aims to connect everything at any time and place
% 
% NTERNET of Things (IoT) 
%
%
%
%The beyond-5G and 6G wireless communication networks are expected to enable ultra-reliable connectivity in high-mobility scenarios. Although conventional orthogonal frequency division multiplexing (OFDM) has shown high spectrum efficiency and robustness under slow fading frequency selective channels, the relative motion between transmitters and receivers in high-mobility scenarios induces severe Doppler spread that inevitably disrupts the sub-carrier orthogonality in conventional OFDM \cite{b1}. This degradation generates inter-carrier interference (ICI), leading to severe performance loss. Therefore, the high-reliable communication over doubly-dispersive channels has become important and challenging for next-generation wireless communication systems.

To overcome the limitations of conventional OFDM, the orthogonal time frequency space (OTFS) modulation is proposed for high-mobility scenarios \cite{b2, b3, b4}. It can effectively utilize the potential delay-Doppler channel diversity for performance improvement and show significant superiority than OFDM under doubly-dispersive channels. The orthogonal chirp division multiplexing (OCDM) \cite{b5} modulation employs chirp signals as sub-carriers and also shows better performance than OFDM. However, the fixed chirp rate in OCDM reduces its robustness against doubly-dispersive channels and limits the performance to particular channel delay-Doppler profiles. To this end, the recently proposed affine frequency division multiplexing (AFDM) modulation \cite{b6, b7, b8, b8.1} maps the transmitted symbols into the discrete affine Fourier (DAF) domain and allows for flexible adjustment of the chirp parameters based on channel spread characteristics, which can provide a complete delay-Doppler channel representation in the DAF domain and achieve the performance advantages similar to those of OTFS. In addition, AFDM has lower pilot guard overhead and enables stronger anti-jamming capability compared to OTFS \cite{b6, b9}, while demonstrating strong robustness in integrated sensing and communication (ISAC) \cite{b10, b11, b12}, which has been thereby considered as a promising solution for next-generation wireless communication systems.

Currently, AFDM has been preliminarily explored across various critical aspects. The work by \cite{b13} proposed a grouped pre-chirp selection method for peak-to-average power ratio (PAPR) reduction. 
The work by \cite{b6} demonstrated that AFDM can achieve full diversity gain based on the optimized chirp parameters under doubly-dispersive channels.
Furthermore, the work by \cite{b8.1} re-optimize the chirp parameters to accommodate the time-scaling characteristics in wideband doubly-dispersive channels and propose the cross domain distributed orthogonal approximate message passing algorithm for low-complexity detection. Then, 
an AFDM-empowered sparse code multiple access (SCMA) system was proposed in \cite{b14} to support massive connectivity in high-mobility scenarios.
In \cite{b15, b16}, the index modulation (IM)-assisted AFDM schemes were proposed to further improve the transmission reliability and energy efficiency of the AFDM systems.
Although the AFDM systems have shown significant superiority in the mentioned works, they mainly depend on the assumption of perfect channel state information (CSI), which has to be estimated in practice.
Therefore, accurate channel estimation is critical to achieve the expected performance of AFDM systems in doubly-dispersive channels.

% 执行信道估计每隔一段时间  

Compared to the rapidly time-varying time-frequency channel response, the original physical channel parameters and DAF domain equivalent channel response
experience a longer stationary time \cite{b16.1, b16.2}, i.e., the channel gain, delay and Doppler shift approximately remain constant over the duration of one or even multiple AFDM frames. Therefore, we can estimate the original physical channel parameters or DAF domain equivalent channel response 
periodically for every few AFDM symbols to track the channel state and prevent channel aging.
However, the duration of an AFDM symbol is always finite due to the low-latency demand in actual internet of vehicles, which cannot provide a sufficient Doppler resolution, leading to the fractional Doppler effect and deteriorating the sparsity of DAF domain equivalent channel matrix.
Furthermore, the delay-Doppler channel spread inevitably introduces interference between pilot and data symbols. These limitations impose serious challenges for accurate channel estimation in AFDM systems.
To mitigate the interference between pilot and data symbols, an embedded pilots structure was designed for AFDM transmission by adding zero-padded symbols as guard intervals to separate the pilot and data symbols \cite{b6, b17}. Then, a threshold-based path detection scheme was proposed in \cite{b17} to extract the DAF domain equivalent channel response. Although it can be achieved with a low complexity, it is sensitive to noise power and only considers the integer Doppler shift. To tackle this issue, an approximated maximum likelihood channel estimation scheme was proposed in \cite{b6} for fractional channel scenarios and achieves better estimation accuracy. However, it requires to traverse possible delay-Doppler combinations, leading to a high computational complexity. Thus, a linear minimum mean squared error (LMMSE) estimator \cite{b18} was proposed for complexity reduction.
Furthermore, a low-complexity diagonal reconstruction channel estimation scheme was proposed for multiple-input multiple-output (MIMO) AFDM systems \cite{b19}, but it only estimates the DAF domain equivalent channel matrix rather than the original physical channel parameters, which limits the estimation accuracy.

To further enhance the channel estimation accuracy, compressed sensing technologies transform the channel estimation into a sparse recovery problem by introducing a virtual delay-Doppler sampling grid, which can effectively exploit the inherent channel sparsity and directly estimate the original delay-Doppler channel response.
The orthogonal matching pursuit (OMP) \cite{b20, b21} algorithms can effectively extract the sparse delay-Doppler channel parameters based on a predefined virtual sampling grid, but it can only estimate the on-grid components, leading to a mismatch measurement matrix for the off-grid component due to fractional channel effects.
Then, the Newtonized OMP (NOMP) was proposed to extract the off-grid component via Newton’s method \cite{b22}, but the performance is limited to the greedy OMP approach.
To overcome the performance limitation of OMP-type estimators, the sparse Bayesian learning (SBL) \cite{b23} algorithm shows better estimation performance by developing a hierarchical hyper-prior to exploit the channel sparsity. However, the original SBL estimators only work for on-grid channel elements and fail to extract the off-grid components.
To tackle the off-grid estimation problem, the multi-resolution grid refinement (MRGR) \cite{b24} is proposed for improving the resolution around promising grid point, but it simultaneously expands the measurement matrix dimension, leading to a significant computational burden. Then, the off-grid SBL (OG-SBL) \cite{b25, b26, b27} extracts off-grid components based on the first-order linear approximation without expanded measurement matrix dimension, but it fails to address the linear approximation error, leading to a significant performance loss. Furthermore, the grid evolution based on grid fission \cite{b27.1} method is proposed for mitigating the off-grid gap by fissioning the grid points and improving local grid resolution, but it also expands the measurement matrix dimension, leading to an increased complexity.
In addition, the SBL-type estimators also involve the covariance matrix inverse operation, leading to a high computational complexity. Although iterative approximation methods \cite{b28, b29, b30} can mitigate the complexity of matrix inversion, there is still a high processing latency due to sequential updating during the iterative process.
Note that the existing channel estimators mentioned above
in AFDM systems either suffer from high complexity and computational latency, or a significant performance loss. Therefore, 
developing an efficient channel estimator is important and challenge to achieve high-precision channel estimation over doubly-dispersive channels.

In this paper, we consider a more practical AFDM system in doubly-dispersive channels with fractional channel effects, and develop an embedded pilot surrounded by guard symbols structure for channel estimation. To overcome the high complexity and performance loss challenge in existing channel estimators, we develop a dynamic grid update strategy instead of applying conventional fixed uniform virtual grid \cite{b20, b21, b22, b23, b24, b25, b26, b27}. We then propose two off-grid SBL channel estimators, i.e., grid refinement SBL (GR-SBL) and grid evolution SBL (GE-SBL), which dynamically adjust the virtual grid for estimation performance improvement. By leveraging the unique sparse structure of the AFDM measurement matrix, we customize a distributed computing scheme to further reduce the complexity of SBL framework, and propose the corresponding distributed GR-SBL (D-GR-SBL) and distributed GE-SBL (D-GE-SBL) estimators, respectively.
The main contributions of this paper are summarized as follows:
\vspace{-3pt}
\begin{itemize}
	\item{Based on the input-output relationship of AFDM transmission with an embedded pilot structure in doubly-dispersive channels, we formulate the channel estimation of AFDM as a sparse signal recovery problem by introducing a virtual sampling grid,
	and develop a SBL framework for AFDM systems to extract the on-grid delay-Doppler components based on a hierarchical hyper-prior distribution.}
	\item{We further develop two off-grid channel estimation methods based on the SBL framework to extract the off-grid Doppler components, termed as GR-SBL and GE-SBL estimators, respectively.
	The GR-SBL estimator can effectively refine the local grid for high-precision estimation, and the GE-SBL estimator can gradually perform grid evolution to mitigate the linear approximation error and improve the channel estimation accuracy.}
	\item{We develop a distributed computing scheme that decomposes the high-dimensional channel estimation model into multiple manageable low-dimensional sub-models for complexity reduction, while leveraging the weak correlation among sub-models to maintain comparable performance to that of the GR-SBL and GE-SBL estimators, termed as D-GR-SBL and D-GE-SBL estimators, respectively. 	
	In addition, the proposed distributed estimators also support parallel computing, which can further reduce the computational latency.}
	\item{The simulation results demonstrate that the proposed estimators outperform the existing channel estimation schemes. The proposed GR-SBL estimator can achieve high-precision estimation with fine search step sizes but with a significant computational burden, while the proposed GE-SBL estimator can achieve a better trade-off between the performance and complexity. In addition, the proposed distributed schemes can significantly reduce the complexity of GR-SBL and GE-SBL estimators while maintaining comparable performance.}
\end{itemize}

The remainder of this paper is organized as follows: Section \uppercase\expandafter{\romannumeral2} introduces the AFDM system model in doubly-dispersive channels. In Section \uppercase\expandafter{\romannumeral3}, we first formulate the AFDM channel estimation model, and then propose GR-SBL and GE-SBL estimators, respectively. The low-complexity distributed channel estimation schemes are proposed in Section \uppercase\expandafter{\romannumeral4}. 
Simulation results are provided in Section \uppercase\expandafter{\romannumeral5} and conclusion is drawn in Section \uppercase\expandafter{\romannumeral6}, respectively. Some detailed proof derivations are provided in the Appendix.

\emph{Notations}: Matrices and vectors are distinguished by bold capital letters and bold lowercase letters. $ \left( \cdot \right)^{*} $, $ \left( \cdot \right)^{\text{T}}$ and $ \left( \cdot \right)^{\text{H}} $ represent the conjugate, transpose and conjugate transpose, respectively. $\mathbb{C}$ denotes the complex number field. $\mathbb{A}$ denotes the normalized constellation set. $\mathbb{E}$ denotes the expectation. $ \boldsymbol{I}_{M \times M} $ and $ \boldsymbol{0}_{M \times 1} $ represent the $M$-dimension identity matrix and all-zeros column vector, respectively. $\text{diag} \{\cdot\}$ denotes the diagonal matrix. $\mathcal{R} (\cdot) $ denotes the real component extraction. $\odot $ denotes the
Kronecker product. $[\cdot]_{N}$ denotes the mod $N$ operation. $\mathcal{CN}  (\cdot)$, $\mathcal{U}  (\cdot)$ and $\Gamma  (\cdot)$ denote the complex Gaussian distribution, uniform distribution and Gamma distribution, respectively. For clarity, we summarize the main system parameter notations of the paper in \textbf{Table \ref{tab1}.}

\begin{table}[t!]\color{black}
	\renewcommand{\arraystretch}{1.5}
	\begin{center}
		\caption{Notations for main system parameters}
		\label{tab1}
		\begin{tabular}{c|c}
			\hline 
			\text{Notations} &  \text{Physical meaning} \\ 
			\hline
			$T$ & Duration of AFDM symbols \\
			$\Delta f$ & Sub-carrier spacing \\
			$N$ & Number of chirp sub-carriers\\
			$\mathcal{M}_p$ & Pilot index set with size $|\mathcal{M}_p|$ \\
			$M_{\text{T}}$ &  Number of received signals for channel estimation \\
			$M_S$ & Virtual sampling grid size \\
			$C$ & Number of groups for distributed computing \\
			$h$ & Complex channel gain \\
			$\ell$ & Normalized delay \\ 
			$k$ & Normalized integer Doppler shift \\
			$\beta$ & Normalized fractional Doppler shift \\
			$\bar{h}$ & Channel gain parameter to be estimated \\
			$\bar{\ell}$ & Virtual delay sampling grid \\
			$\bar{k}$ & Virtual Doppler sampling grid \\
			$\boldsymbol{y}_{\text{T}}$ & Received signals for channel estimation\\
			$\boldsymbol{y}_{\text{T},c}$ & $c$-th group received signals for distributed computing \\ 
			$\boldsymbol{\Phi}$ & Measurement matrix for channel estimation  \\ 
			$\boldsymbol{\Phi}_c$ & $c$-th group measurement matrix for distributed computing \\
			\hline
		\end{tabular}
	\end{center}
\end{table}

\section{AFDM System Model}
\begin{figure*}[t!]
	\centering
	\includegraphics[scale=0.6]{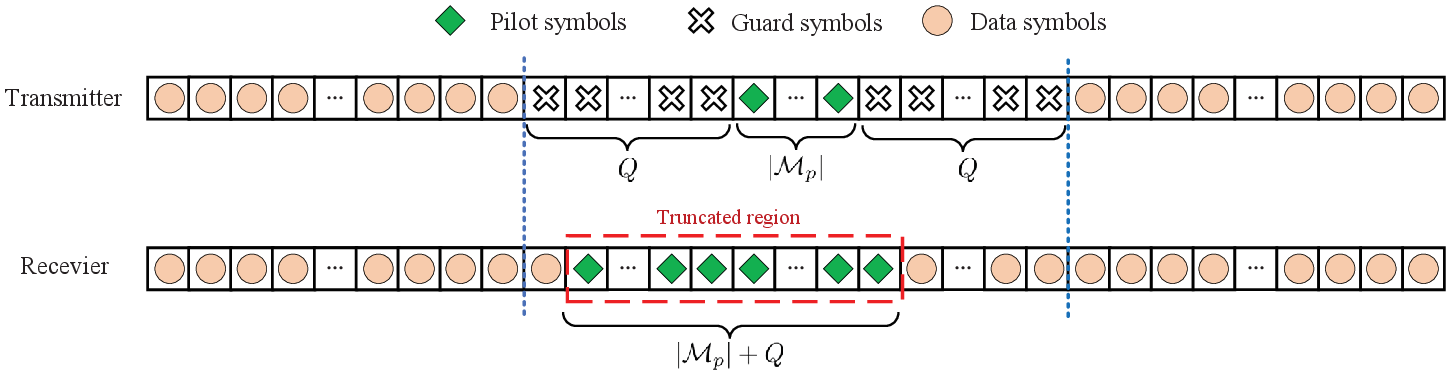}
	\caption{Pilot and data pattern for channel estimation in AFDM systems.}
	\label{Fig_1}
	\vspace{-15pt}
\end{figure*} 
The AFDM modulation has shown strong performance advantage in high-mobility scenarios, which is considered as a promising waveform to enable high-reliable connectivity in internet of vehicles. 
Without loss of generality, we consider an $N$ chirp sub-carrier AFDM system with duration $T$ and bandwidth $N \varDelta f$, where $T = 1 / \varDelta f$. According to \cite{b6}, the AFDM system can provide a complete delay-Doppler channel representation in the DAF domain based on the optimized chirp parameters, which motivates us to formulate the embedded pilot pattern in the DAF domain, as shown in Fig~.\ref{Fig_1}. 
Specifically, the pilot symbols $\boldsymbol{x}_p \in \mathbb{C}^{| \mathcal{M}_{p} | \times 1}$ are first placed in the DAF domain, where $\mathcal{M}_{p} $ is the pilot index set with size $| \mathcal{M}_{p} |$. Then, the zero-padded symbols are placed surrounding the pilot symbols as guard intervals with size $2Q$ for mitigating the interference between pilot symbols and data symbols, where $Q = (\ell_{max} + 1)(2k_{max} + 2  N_v + 1) - 1$, $\ell_{max}$ and $k_{max}$ are the maximum normalized delay and maximum normalized Doppler shift, $N_v$ is a non-negative integer that mitigates the interference spread caused by the fractional Doppler effect and aids the design of chirp parameters for achieving high channel diversity gain.
The modulated data symbols $\boldsymbol{x}_d \in \mathbb{C}^{(N-2Q-|\mathcal{M}_p|) \times 1} $ from a constellation set  $\mathbb{A}$ (e.g., PSK, QAM symbol sets) are placed in the remaining positions. Therefore, the AFDM transmission frame can be given by
\begin{align}\label{1}
	x[m]  = 
	\begin{cases}
		x_{{p}} [m], & m \in \mathcal{M}_{p} \\ 
		0, &  \min\{\mathcal{M}_{p}\} - Q \leq m \leq \max\{\mathcal{M}_{p}\} + Q\\ & m  \notin \mathcal{M}_{p} \\
		x_{{d}} [m], & \text{otherwise} 
	\end{cases}.
\end{align}

\vspace{-6pt}
The AFDM transmitted symbols $x[m]$  are firstly mapped into DAF domain and then transform to the time domain by inverse DAF transform, i.e.,
\begin{align}\label{2}
	s [n] = \frac{1}{\sqrt{N}} \sum_{m=0}^{N-1} x[m] e^{j 2 \pi (c_1 n^2 + nm/N + c_2 m^2)},
\end{align}
where $c_1$ and $c_2$ are the AFDM chirp parameters. 
To combat the interference between adjacent AFDM transmission blocks from multipath propagation and keep the continuous periodicity of AFDM symbols, we add the chirp-periodic prefix (CPP) no shorter than the maximal channel delay spread $L_{cpp}$ as
\vspace{-3pt}
\begin{align}\label{3}
	s [n] = s [N + n] e^{-j2 \pi c_1 (N^2 + 2N n)}, n = -L_{cpp}, ..., -1.
\end{align}  

The doubly-dispersive channels can be described based on the time-delay channel response $h[n, \ell]$ as
\vspace{-3pt}
\begin{align}\label{4}
	h[n,\ell] = \sum_{{i_p}=0}^{P-1} h_{i_p} e^{-j 2 \pi (k_{i_p} + \beta_{i_p}) n / N} \delta [\ell - \ell_{i_p}],
\end{align}
where $P$ is the number of propagation paths. 
$h_{i_p}$, $\ell_{i_p} \in [0, \ell_{max}]$, $k_{i_p} \in [-k_{max}, k_{max}]$ and $\beta_{i_p} \in \left( -1/2, 1/2\right] $ represent the ${i_p}$-th path complex gain, normalized delay, normalized integer and fractional Doppler shift, respectively.\footnote{\label{n1}Note that we do not need to consider fractional delays due to the sufficient supporting of the sampling time resolution in typical wide-band systems \cite{b6}.}

The time domain signal $s[n]$ is transmitted over the doubly-dispersive channel $h [n, \ell]$. At the receiver, after removing the CPP, the received signal $r[n]$ can be expressed as
\vspace{-3pt}
\begin{align}\nonumber
	r [n] & = \sum_{\ell=0}^{\infty} s [n - \ell]_{N} h [n, \ell] + \tilde{\omega} [n] \label{5} \\
	& = \sum_{{i_p}=0}^{P-1} s[n- \ell_{i_p}]_{N} e^{-j 2 \pi (k_{i_p} + \beta_{i_p}) n / N} + \tilde{\omega} [n],
\end{align}
where $\tilde{\omega}[n]$ is additive white Gaussian noise (AWGN) with zero-mean and variance ${\gamma}^{-1}$.

Finally, the DAF domain received signal $y [\tilde{m}]$ can be obtained by the DAF transform to $r[n]$,
\begin{align}\label{6}
	y [\tilde{m}] = \frac{1}{\sqrt{N}} \sum_{n=0}^{N-1} r[n] e^{-j 2 \pi (c_1 n^2 + \tilde{m}n/N + c_2 \tilde{m}^2)}.
\end{align}

By substituting \eqref{2} and \eqref{5} into \eqref{6}, the input-output relationship of AFDM systems in doubly-dispersive channels can be given by
\begin{align}\label{7}
	\boldsymbol{y} = \boldsymbol{H} \boldsymbol{x} + \boldsymbol{\omega},
\end{align}
where $\boldsymbol{\omega}$ is the equivalent noise in the DAF domain with same distribution as $\boldsymbol{\tilde{\omega}}$, and $\boldsymbol{H}$ is the DAF domain equivalent channel matrix, which is given by
\begin{align}\label{8}
	H [\tilde{m},m] \!=\! \frac{1}{N} \sum_{{i_p}=0}^{P-1} h_{i_p} e^{j 2 \pi \!\left[\! c_1 \ell_{{i_p}}^2 \!-\! m \ell_{{i_p}}/N \!+\! c_2 (m^2 \!-\! \tilde{m}^2) \!\right]\! } \mathcal{F}_{i_p} \!\left(\! \tilde{m} , m \!\right)\!,
\end{align}
with
\begin{align}\nonumber
	\mathcal{F}_{i_p} \left( \tilde{m} , m \right) & = \sum_{n=0}^{N-1} e^{ - j \frac{2 \pi}{N} \left[ (\tilde{m} - m + 2Nc_1 \ell_{i_p} + k_{i_p} + \beta_{i_p}) n \right] } \\ \label{9}
	& = \frac{e^{ - j 2 \pi \left[ (\tilde{m} - m + 2Nc_1 \ell_{i_p} + k_{i_p} + \beta_{i_p}) \right] } - 1}{e^{ - j \frac{2 \pi}{N} \left[ (\tilde{m} - m + 2Nc_1 \ell_{i_p} + k_{i_p} + \beta_{i_p}) \right] } - 1}.
\end{align}

According to \cite{b6}, the AFDM chirp parameters $c_1$ and $c_2$ can be optimized based on the channel spread characteristics for improving the diversity gain and system performance, which can be given by
\begin{align}\label{10}
	c_1 = \frac{2 k_{max} + 2 N_v + 1}{ 2 N }
\end{align}
and $c_2$ is an arbitrary irrational number. In addition, the number of transmitted symbols $N$ should satisfy $ N > (2 k_{max} + 2 N_v + 1)(\ell_{max} + 1)  $ for full channel diversity achievement.

\section{Proposed Channel Estimation with Sparse Bayesian Learning }
In this section, we first formulate the channel estimation of AFDM systems as a sparse signal recovery problem. Then, we develop a SBL framework for AFDM systems and propose two off-grid channel estimation methods for extracting the delay-Doppler channel parameters, i.e., GR-SBL and GE-SBL estimators. 

\subsection{Channel Estimation Model}

%\begin{figure*}[t!]
%	\centering
%	\includegraphics[scale=0.68]{Pilot.eps}
%	\caption{Pilot and data pattern for channel estimation in AFDM system.}
%	\label{Fig_1}
%\end{figure*} 

%Based on the optimized AFDM chirp parameter in \eqref{9}, it can provide a complete delay-Doppler channel representation in DAF domain, which motivates us to establish the channel estimation model in DAF domain \cite{c6}. Specifically, the pilot symbols $\boldsymbol{x}_0$ surrounded by guard symbols are placed in DAF domain, where the guard symbols are used to mitigate the interference between pilot symbols $\boldsymbol{x}_0$ and data symbols $\boldsymbol{x}_d$, which can be described in Fig .\ref{Fig_1}. The AFDM transmission frame can be given by
%\begin{align}\label{10}
%	x[m]  = 
%	\begin{cases}
%		x_{{0}} [m], & m \in \mathcal{M}_{0} \\ 
%		0, &  \min\{\mathcal{M}_{0}\} - Q \leq m \leq \max\{\mathcal{M}_{0}\} + Q\\ & m  \notin \mathcal{M}_{0} \\
%		x_{{d}} [m], & \text{otherwise} 
%	\end{cases},
%\end{align}
%where $Q = (\ell_{max} + 1)(2k_{max} + 2  N_v + 1) - 1$, $\mathcal{M}_{0} $ is the pilot index set with size $| \mathcal{M}_{0} |$.  

\begin{figure}[htpt]
	\centering
	\includegraphics[scale=0.55]{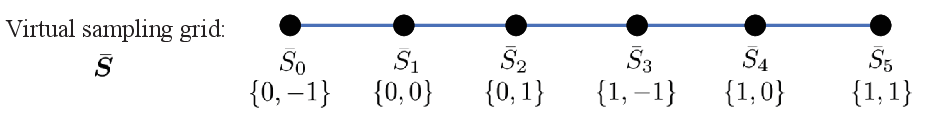}
	\caption{An example structure of virtual sampling grid $\boldsymbol{\bar{S}}$ with $\boldsymbol{\tilde{\ell}}  = [0, 1]^{\text{T}}$ and $\boldsymbol{\tilde{k}} = [-1, 0, 1]^{\text{T}}$.}
	\label{Fig_2.1}
\end{figure}

According to the input-output relationship of the AFDM system in \eqref{7}, the received pilot signals are primarily concentrated within a certain region of the DAF domain. This enables to consider a truncated received pilot signal $ \boldsymbol{y}_{\text{T}} = \{ y[\tilde{m}] | \tilde{m} \in \left[ \min\{\mathcal{M}_p\} \!-\! Q \!+\! k_{max} \!+\! N_v, \max\{\mathcal{M}_{p}\} \!+\! k_{max} \!+\! N_v \right] \}$ with size $M_{\text{T}} \!=\! | \mathcal{M}_{p} | \!+\! Q$ for channel estimation, as shown in Fig.~\ref{Fig_1}. For brevity, we let $S_{i_p} = \{\ell_{i_p}, k_{i_p} + \beta_{i_p}\}$ include the delay and Doppler of the $i_p$-th path, $i_p = 0, 1, ..., P-1$. According to \eqref{7},
the input-output relationship of the truncated received pilot signal $\boldsymbol{y}_{\text{T}}$ can be rewritten as
\begin{align}\label{11}
	\boldsymbol{y}_{\text{T}} = \boldsymbol{\Phi} (\boldsymbol{{S}})    \boldsymbol{{h}} + \boldsymbol{\omega}_{\text{T}},
\end{align}
where $\boldsymbol{h} = [h_0, h_1, ..., h_{i_p}, ..., h_{P-1}]^{\text{T}} \in \mathbb{C}^{P \times 1}$,
$\boldsymbol{y}_{\text{T}} \in \mathbb{C}^{M_{\text{T}} \times 1}$, and $\boldsymbol{\omega}_{\text{T}} \in \mathbb{C}^{M_{\text{T}} \times 1}$ is the measurement noise. $\boldsymbol{\Phi} (\boldsymbol{{S}}) \in \mathbb{C}^{M_{\text{T}} \times P}  $ is the measurement matrix given by
\begin{align}\label{12}
	\boldsymbol{\Phi} (\boldsymbol{{S}}) = \left[ \boldsymbol{\phi} ({S}_0), \boldsymbol{\phi} ({S}_1) ..., \boldsymbol{\phi} ({S}_{i_p}), ..., \boldsymbol{\phi} ({S}_{P-1}) \right],
\end{align}
where $ \boldsymbol{\phi} ({S}_{i_p}) \in \mathbb{C}^{M_{\text{T} }\times 1}$ and the corresponding $\bar{m}$-th entry is given by
\begin{align}\nonumber
	{\phi}_{\bar{m}} ({S}_{i_p})  & = \frac{1}{N} \sum_{m \in \mathcal{M}_{p}} x[m] e^{j 2 \pi \left[ c_1 {\ell}_{i_p}^2 - {m} {\ell}_{i_p} /N + c_2 ({m}^2 - \tilde{m}^2) \right] } \\ \label{13}
	& \times \frac{e^{ - j 2 \pi \left[ (\tilde{m} - m + 2Nc_1 {\ell}_{i_p} + {k}_{i_p} + \beta_{i_p}) \right] } - 1}{e^{ - j \frac{2 \pi}{N} \left[ ( \tilde{m} - m + 2Nc_1 {\ell}_{i_p} + {k}_{i_p} + \beta_{i_p}) \right] } - 1}
\end{align}
with $\tilde{m} \!=\! \min\{\mathcal{M}_p\} - Q + k_{max} + N_v + \bar{m}$, $\bar{m} \!=\! 0, 1, ..., M_{\text{T}}-1$. According to \eqref{12} and \eqref{13}, the measurement matrix $\boldsymbol{\Phi} (\boldsymbol{{S}})$ is characterized based on unknown channel information $\ell_{i_p}$, $k_{i_p} + \beta_{i_p}$ and $P$. Therefore, it is infeasible to directly estimate the channel parameters based on \eqref{11}. To overcome this challenge, we transform the channel estimation into a sparse signal recovery problem by establishing a virtual sampling grid. Specifically, we first 
perform the virtual sampling along the delay and 
Doppler dimension in the range of $\left[ 0, \ell_{max}\right] $ and $\left[ -k_{max}-1, k_{max}+1\right] $  with delay resolution $r_{\tau} = \frac{\ell_{max} }{M_{\tau}-1}$ and Doppler resolution $r_{\nu} = \frac{2k_{max} + 2}{M_{\nu}-1}$. Then, the 
virtual sampling delay and Doppler vectors can be given by $\boldsymbol{\tilde{\ell}}  = [\tilde{\ell}_0, \tilde{\ell}_1, ..., \tilde{\ell}_{M_{\tau}-1}]^{\text{T}}$ and $\boldsymbol{\tilde{k}} = [\tilde{k}_0, \tilde{k}_1, ..., \tilde{k}_{M_{\nu}-1}]^{\text{T}}$ with $\tilde{\ell}_a = a r_{\tau}, a = 0, 1, ..., M_{\tau}-1$ and $\tilde{k}_b = b r_{\nu} - k_{max} - 1, b = 0, 1, ..., M_{\nu}-1$. Furthermore, we define the virtual sampling grid as $\bar{S}_i = \{\bar{\ell}_i, \bar{k}_i\}, i = 0, 1, ..., M_S - 1$ with $M_S = M_{\tau} M_{\nu}$, which includes all possible combinations of virtual delay and Doppler, i.e., for any ${\tilde{\ell}_a, \tilde{k}_b}$, we always have the corresponding virtual sampling grid $\bar{S}_i$ that satisfies  $\bar{\ell}_i = \tilde{\ell}_a$, $\bar{k}_i = \tilde{k}_b$ with $a = \lfloor \frac{i}{M_{\nu}} \rfloor$, $b = i - a M_{\nu}$. For more clarity, we take the virtual sampling delay vector $\boldsymbol{\tilde{\ell}}  = [0, 1]^{\text{T}}$ and virtual sampling Doppler vector $\boldsymbol{\tilde{k}} = [-1, 0, 1]^{\text{T}}$
as an example to illustrate the structure of virtual sampling grid $\boldsymbol{\bar{S}}$, as shown in Fig. \ref{Fig_2.1}. 

Then, the channel estimation model of the AFDM system can be established as
\begin{align}\label{14}
	\boldsymbol{y}_{\text{T}} = \boldsymbol{\Phi} (\boldsymbol{\bar{S}})    \boldsymbol{\bar{h}} + \boldsymbol{\omega}_{\text{T}},
\end{align}
where $\boldsymbol{\Phi} (\boldsymbol{\bar{S}}) \in \mathbb{C}^{M_{\text{T}} \times M_S}  $ is the measurement matrix based on virtual sampling grid $\boldsymbol{\bar{S}}$, which can be given by
\begin{align}\label{15}
	\boldsymbol{\Phi} (\boldsymbol{\bar{S}}) = \left[ \boldsymbol{\phi} (\bar{S}_0), \boldsymbol{\phi} (\bar{S}_1) ..., \boldsymbol{\phi} (\bar{S}_i), ..., \boldsymbol{\phi} (\bar{S}_{M_S-1}) \right].
\end{align}
Here, $\boldsymbol{\bar{h}} = [\bar{h}_0, \bar{h}_1, ..., \bar{h}_i, ..., \bar{h}_{M_S-1}]^{\text{T}} \in \mathbb{C}^{M_S \times 1}$ is the channel vector to be estimated. Note that although there are $M_S$ elements in $\boldsymbol{\bar{h}}$, only $P$ elements are non-zeros if $P$ separable propagation paths fall into the virtual sampling grid with $P \ll M_S $, which forms an obvious sparse structure. According to the compressed sensing techniques \cite{b25}, for a sparse vector $\boldsymbol{\bar{h}}$ with length $M_S$, there are at most 
$\bar{P} = \lfloor \frac{M_{\text{T}}}{\ln(M_S)} \rfloor$ non-zeros element can be recovered based on $M_{\text{T}}$ measurements. Therefore, we can initialize the number of paths as $\bar{P}$ and the channel estimation model in \eqref{11} is transformed into a sparse signal recovery problem in \eqref{14}. 

\subsection{Sparse Bayesian Learning}
In this sub-section, we develop a SBL framework for channel estimation in AFDM systems, which can effectively exploit the potential sparsity to enhance the channel estimation accuracy. Specifically, we first  formulate the channel vector $ \boldsymbol{\bar{h}} $ as a hierarchical hyper-prior Laplace distribution, i.e., 
\begin{align}\label{16}
	\Pr (\boldsymbol{\bar{h}} \left| \boldsymbol{\alpha} \right. ) & = \mathcal{CN} (\boldsymbol{\bar{h}} \left| \boldsymbol{0}_{M_S \times 1}, \boldsymbol{\Lambda}\right. ), \\ \label{17}
	\Pr (\boldsymbol{\alpha} \left| \rho \right. ) & = \prod_{i=0}^{M_S-1} \Gamma (\alpha_i | 1, \rho),
\end{align}
where $\boldsymbol{\Lambda} = \text{diag} \{ \boldsymbol{\alpha} \}$, $\boldsymbol{\alpha} = [\alpha_0, ..., \alpha_{M_S - 1}]^{\text{T}}$ is the variance vector of $\boldsymbol{\bar{h}}$, $\rho$ is a root parameter of $\boldsymbol{\alpha}$. The measurement noise is also unknown and we assume that the noise vector $\boldsymbol{\omega}_{\text{T}}$ follows a complex Gaussian distribution, i.e.,
\begin{align}\label{18}
	\Pr (\boldsymbol{\omega}_{\text{T}} \left| \gamma \right. ) = \mathcal{CN} (\boldsymbol{\omega}_{\text{T}} \left| \boldsymbol{0}_{M_{\text{T}} \times 1}, \gamma^{-1} \boldsymbol{I}_{M_{\text{T}}}\right. ),
\end{align}
where $\gamma$ denotes the noise precision and it further follows a Gamma distribution,
\begin{align}\label{19}
	\Pr (\gamma \left| c, d\right.) =  \Gamma ( \gamma | c, d )
\end{align}
with root parameters $c$ and $d$. Moreover, the Doppler component may not fall into the predefined virtual sampling grid due to insufficient Doppler resolution. Then, the off-grid Doppler component $\bar{\beta}_i$ corresponding to the $i$-th virtual sampling grid is assumed to follow a uniform distribution, i.e.,
\begin{align}\label{20}
	{\bar{\beta}}_i = \mathcal{U} \left[ -\frac{1}{2} r_{\nu}, \frac{1}{2} r_{\nu} \right].
\end{align}

Before implementing the SBL framework, we first formulate the measurement matrix $\boldsymbol{\Phi} (\boldsymbol{\bar{S}}^{(t)}) $ based on the virtual sampling grid $\boldsymbol{\bar{S}}^{(t)}  $ by \eqref{15} and set the initial off-grid Doppler component $\boldsymbol{\bar{\beta}}^{(t)} = \boldsymbol{0}_{M_S \times 1} $ at the beginning of the $(t)$-th iteration. 
The conditional posteriori distribution $\Pr (\boldsymbol{\bar{h}} |\boldsymbol{y}_{\text{T}};  \boldsymbol{\alpha}^{(t)},  \gamma^{(t)}, \boldsymbol{\bar{S}}^{(t)}) $ with the given parameters $\left( \boldsymbol{\alpha}^{(t)},  \gamma^{(t)}, \boldsymbol{\bar{S}}^{(t)} \right)$  can be expressed as
\begin{align}\label{21.0}
	\Pr (\boldsymbol{\bar{h}} |\boldsymbol{y}_{\text{T}};  \boldsymbol{\alpha}^{(t)},  \gamma^{(t)}, \boldsymbol{\bar{S}}^{(t)}) \propto 
	\Pr ( \boldsymbol{y}_{\text{T}} |\boldsymbol{\bar{h}};  \gamma^{(t)}, \boldsymbol{\bar{S}}^{(t)}) \Pr ( \boldsymbol{\bar{h}}; \boldsymbol{\alpha}^{(t)}),
\end{align}
where the priori distribution $\Pr ( \boldsymbol{\bar{h}}; \boldsymbol{\alpha}^{(t)})$ is given by \eqref{16}, and the conditional likelihood function $\Pr ( \boldsymbol{y}_{\text{T}} |\boldsymbol{\bar{h}};  \gamma^{(t)}, \boldsymbol{\bar{S}}^{(t)})$ is given by
\begin{align}\label{21.1}
	\Pr ( \boldsymbol{y}_{\text{T}} |\boldsymbol{\bar{h}};  \gamma^{(t)}, \boldsymbol{\bar{S}}^{(t)}) = \mathcal{CN} \left( \boldsymbol{y}_{\text{T}} | \boldsymbol{\Phi} (\boldsymbol{\bar{S}}^{(t)})    \boldsymbol{\bar{h}},  \left( \gamma^{(t)}\right)^{-1}  \boldsymbol{I}_{M_{\text{T}}}\right).
\end{align}

Combining \eqref{16}, \eqref{21.0} and \eqref{21.1}, 
the conditional posteriori distribution $ \Pr (\boldsymbol{\bar{h}} |\boldsymbol{y}_{\text{T}};   \boldsymbol{\alpha}^{(t)},  \gamma^{(t)}, \boldsymbol{\bar{S}}^{(t)} ) $ can be expressed as
\begin{align}\label{21}
	\Pr (\boldsymbol{\bar{h}} |\boldsymbol{y}_{\text{T}};  \boldsymbol{\alpha}^{(t)}, \gamma^{(t)}, \boldsymbol{\bar{S}}^{(t)}) = \mathcal{CN} (\boldsymbol{\bar{h}} | \boldsymbol{\mu}^{(t)}, \boldsymbol{\Sigma}^{(t)}),
\end{align}
where the conditional posteriori variance $\boldsymbol{\Sigma}^{(t)}$ and the mean $\boldsymbol{\mu}^{(t)}$ are given by
\begin{align}\nonumber
	&\boldsymbol{{\Sigma}}^{(t)}  = \boldsymbol{\Lambda}^{(t)}  -  \boldsymbol{\Lambda}^{(t)}  \boldsymbol{{\Phi}}^{\text{H}} (\boldsymbol{\bar{S}}^{(t)}) \\ \label{22}
	& \times   \left( \left( \gamma^{(t)} \right)^{-1} \boldsymbol{I}_{M_{\text{T}}} + \boldsymbol{{\Phi}} (\boldsymbol{\bar{S}}^{(t)}) \boldsymbol{\Lambda}^{(t)} \boldsymbol{{\Phi}}^{\text{H}} (\boldsymbol{\bar{S}}^{(t)})   \right)^{-1} \boldsymbol{{\Phi}} (\boldsymbol{\bar{S}}^{(t)}) \boldsymbol{\Lambda}^{(t)}, \\ \label{23}
	& \boldsymbol{{\mu}}^{(t)} =  \gamma^{(t)} \boldsymbol{{\Sigma}}^{(t)} \boldsymbol{{\Phi}}^{\text{H}} (\boldsymbol{\bar{S}}^{(t)}) \boldsymbol{y}_{\text{T}}. 
\end{align}
%\begin{align}\nonumber
%	&\boldsymbol{{\Sigma}}^{(t)}  = \boldsymbol{\Lambda}^{(t)}  - \\ \label{22} & \boldsymbol{\Lambda}^{(t)}  \boldsymbol{{\Phi}}^{\text{H}} (\boldsymbol{\bar{S}}^{(t)}\!)  \!\left(\! \gamma^{(t)} \boldsymbol{I}_{M_{\text{T}}} \!\!+\! \boldsymbol{{\Phi}} (\boldsymbol{\bar{S}}^{(t)}\!)\! \boldsymbol{\Lambda}^{(t)} \boldsymbol{{\Phi}}^{\text{H}} (\boldsymbol{\bar{S}}^{(t)}\!)\!   \right)^{\!-1}\!\! \boldsymbol{{\Phi}} (\boldsymbol{\bar{S}}^{(t)}\!)\! \boldsymbol{\Lambda}^{(t)}, \\ \label{23}
%	& \boldsymbol{{\mu}}^{(t)} =  \gamma^{(t)} \boldsymbol{{\Sigma}}^{(t)} \boldsymbol{{\Phi}}^{\text{H}} (\boldsymbol{\bar{S}}^{(t)}) \boldsymbol{y}_{\text{T}}. 
%\end{align}

Then, the hyper-parameters $\boldsymbol{{\alpha}}^{(t+1)}$ and ${\gamma}^{(t+1)}$ can be updated to maximize the logarithmic joint distribution 
$ \ln \left[ \Pr (\boldsymbol{y}_{\text{T}}, \boldsymbol{\bar{h}}, \boldsymbol{\alpha}, \gamma ) \right]  $ given the conditional posteriori distribution $ \Pr (\boldsymbol{\bar{h}} |\boldsymbol{y}_{\text{T}}; \boldsymbol{\alpha}^{(t)}, \gamma^{(t)}, \boldsymbol{\bar{S}}^{(t)}) $
based on the expectation-maximization (EM) algorithm \cite{b25}. 
In the E-step, the objective function $Q \left( \boldsymbol{\alpha}, {\gamma} \left|  \boldsymbol{\alpha}^{(t)}, {\gamma}^{(t)}, \boldsymbol{\bar{S}}^{(t)} \right.  \right)$ of logarithmic joint distribution $ \ln \left[ \Pr (\boldsymbol{y}_{\text{T}}, \boldsymbol{\bar{h}}, \boldsymbol{\alpha}, \gamma ) \right] $ by averaging on the
conditional posteriori distribution $\Pr (\boldsymbol{\bar{h}} |\boldsymbol{y}_{\text{T}}; \boldsymbol{\alpha}^{(t)}, \gamma^{(t)}, \boldsymbol{\bar{S}}^{(t)} )$ is given by
\begin{align}\nonumber
	& Q \left(  \boldsymbol{\alpha}, {\gamma} \left| \boldsymbol{\alpha}^{(t)}, {\gamma}^{(t)}, \boldsymbol{\bar{S}}^{(t)} \right.  \right) \\ \label{24}
	& ~~~~~~~ = \mathbb{E}_{\Pr ( \boldsymbol{\bar{h}} |\boldsymbol{y}_{\text{T}}; \boldsymbol{\alpha}^{(t)}, \gamma^{(t)}, \boldsymbol{\bar{S}}^{(t)})}  \left\lbrace  \ln \left[  \Pr (\boldsymbol{y}_{\text{T}}, \boldsymbol{\bar{h}}, \boldsymbol{\alpha}, \gamma ) \right]  \right\rbrace.
\end{align}

In the M-step, the hyper-parameters $\boldsymbol{{\alpha}}^{(t+1)}$ and ${\gamma}^{(t+1)}$ can be updated by maximizing the objective function $Q \left( \boldsymbol{\alpha}, {\gamma} \left| \boldsymbol{\alpha}^{(t)}, {\gamma}^{(t)}, \boldsymbol{\bar{S}}^{(t)} \right.  \right)$, i.e.,
\begin{align}\label{25}
	\left( \boldsymbol{{\alpha}}^{(t+1)}, {\gamma}^{(t+1)}\right)   \!=\! \arg\max_{\boldsymbol{\alpha}, \gamma} Q \left( \boldsymbol{\alpha}, {\gamma} \left| \boldsymbol{\alpha}^{(t)}, {\gamma}^{(t)}, \boldsymbol{\bar{S}}^{(t)} \right.  \right),
\end{align}
which can be given by \cite{b23}
\begin{align}\label{26}
	{\alpha}_i^{(t+1)} & = \frac{\sqrt{1 + 4 \rho \left( \left| \mu_i^{(t)}\right|^2 + \Sigma_{i,i}^{(t)}\right) } - 1}{2 \rho}, \\ \label{27}
	{\gamma}^{(t+1)} & = \frac{c - 1 + M_{\text{T}}}{ d + \mathcal{E}_{\gamma}^{(t)}},
\end{align}
where
\begin{align}\label{28}
	\mathcal{E}_{\gamma}^{(t)} \!=\! \left| \left|  \boldsymbol{y}_{\text{T}} \!-\! \boldsymbol{\Phi} (\boldsymbol{\bar{S}}^{(t)}) \boldsymbol{\mu}^{(t)} \right| \right|^2 \!+\! \left( \gamma^{(t)} \right)^{-1} \sum_{i=0}^{M_S-1} \left( 1 \!-\! \frac{\Sigma_{i,i}^{(t)}}{ \alpha_i^{(t)}}  \right) .
\end{align}

We then extract $\bar{P}$ maximum elements from $\boldsymbol{{\alpha}}^{(t+1)}$ and the corresponding index set is given by $\mathcal{P}$.
Therefore, the on-grid delay and Doppler components can be preliminarily estimated as $ \bar{S}_p^{(t)} = \{\bar{\ell}_p^{(t)}, \bar{k}_p^{(t)}\}$ with $p \in \mathcal{P}$. For the original SBL algorithm \cite{b23}, it employs a fixed virtual sampling grid and $\boldsymbol{\bar{S}}^{(t+1)}$ can be updated as $\boldsymbol{\bar{S}}^{(t+1)} = \boldsymbol{\bar{S}}^{(t)}$. In addition, the original SBL fails to estimate the off-grid components $\boldsymbol{\bar{\beta}}$. 
To mitigate the estimation error caused by off-grid elements, the high-resolution predefined virtual grid is usually adopted to improve the estimation performance \cite{b24}, but it simultaneously expands the dimension of the measurement matrix, leading to a significant computational burden. Although the OG-SBL algorithm \cite{b25,b26,b27} can also extract the off-grid components based on first-order linear approximation, it employs a fixed virtual sampling grid and fails to address the linear approximation error, leading to a significant performance loss. To tackle these limitations and improve the estimation accuracy, we develop a dynamic grid update strategy and propose two off-grid channel estimation methods, i.e., GR-SBL and GE-SBL to estimate the off-grid Doppler
components in the subsequent sub-sections.

\subsection{GR-SBL for Off-Grid Doppler Component Estimation}
\begin{figure}[t!]
	\centering
	\includegraphics[scale=0.55]{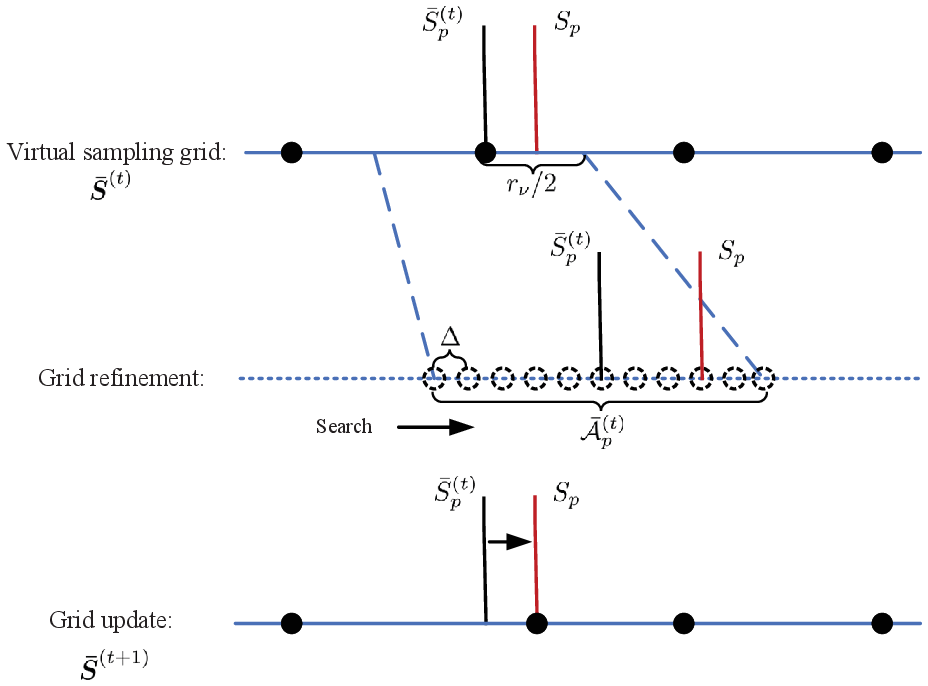}
	\caption{GR-SBL for off-grid Doppler component estimation.}
	\label{Fig_2}
\end{figure} 

In this sub-section, we develop a GR-SBL estimator for extracting off-grid Doppler component in AFDM systems. By conducting a local grid refinement search around the on-grid component $\bar{S}_p^{(t)}, p \in \mathcal{P}$  preliminarily estimated via SBL, the proposed GR-SBL estimator can effectively increase the resolution around the estimated on-grid components and prevent the expansion of measurement matrix dimension, avoiding the high computational burden. The grid refinement process can be described in Fig. \ref{Fig_2}, where the red line denotes the real Doppler values.

According to the maximum likelihood (ML) criterion \cite{b31, b32, b33}, the conditional likelihood function  $\Pr \left(\boldsymbol{y}_{\mathrm{T}} | \boldsymbol{\alpha}^{(t)}, \boldsymbol{\bar{S}}^{(t)}; \gamma^{(t)}\right)$ based on hyper-parameters $\boldsymbol{\alpha}^{(t)}$ and virtual sampling grid $\boldsymbol{\bar{S}}^{(t)}$ can be expressed as 
\begin{align}\nonumber
	& \Pr \left(\boldsymbol{y}_{\mathrm{T}} | \boldsymbol{\alpha}^{(t)}, \boldsymbol{\bar{S}}^{(t)}; \gamma^{(t)}\right) \\ \nonumber
	& = \int_{-\infty}^{+\infty} \Pr \left(\boldsymbol{y}_{\mathrm{T}}, \boldsymbol{\bar{h}} | \boldsymbol{\alpha}^{(t)}, \boldsymbol{\bar{S}}^{(t)}; \gamma^{(t)}\right) d \boldsymbol{\bar{h}} \\ \nonumber
	& =  \int_{-\infty}^{+\infty} \Pr \left(\boldsymbol{y}_{\mathrm{T}} | {\boldsymbol{\bar{h}}}, \boldsymbol{\bar{S}}^{(t)};  \gamma^{(t)} \right) \Pr \left( {\boldsymbol{\bar{h}}} | \boldsymbol{\alpha}^{(t)} \right)  {d} {\boldsymbol{\bar{h}}} \\ \nonumber
	& = \int_{-\infty}^{+\infty}  \!\!\!\!\! \mathcal{CN} \!\left(\!\! \boldsymbol{y}_{\text{T}} | \boldsymbol{\Phi} (\boldsymbol{\bar{S}}^{(t)})    \boldsymbol{\bar{h}},  \!\left(\! \gamma^{\!(t)}\!\right)^{-1} \!\!\!\! \boldsymbol{I}_{M_{\text{T}}} \!\!\right) \!  \mathcal{CN} (\boldsymbol{\bar{h}} \left| \boldsymbol{0}_{M_S \!\times \! 1},\! \boldsymbol{\Lambda}^{\!(t)}\right. \!)  {d} {\boldsymbol{\bar{h}}} \\
	& = \left( \pi^{M_{\text{T}}} |\boldsymbol{C}|\right)^{-1} \exp \left[ - \operatorname{tr} \left(  \boldsymbol{C}^{-1}  \boldsymbol{R}_y   \right)  \right],  
\end{align}
where $\boldsymbol{R}_y \!=\! \boldsymbol{y}_{\text{T}} \boldsymbol{y}_{\text{T}}^{\text{H}}$, $\boldsymbol{C} \!=\!  {\boldsymbol{\Phi}} (\boldsymbol{\bar{S}}^{(t)}) \boldsymbol{\Lambda}^{(t)} {\boldsymbol{\Phi}}^{\mathrm{H}} (\boldsymbol{\bar{S}}^{(t)}) \!+\! \left( \gamma^{(t)} \right)^{-1}  \boldsymbol{I}_{M_{\text{T}}} $. 
To extract the off-grid Doppler components, the logarithmic objective function ${\mathcal{L}} \left( \boldsymbol{\alpha}^{(t)}, \boldsymbol{\bar{S}}^{(t)}\right)$ based on virtual grid $\boldsymbol{\bar{S}}^{(t)}$ and corresponding hyper-parameters $\boldsymbol{\alpha}^{(t)}$ is given by 
\begin{align}\nonumber
	{\mathcal{L}} \left( \boldsymbol{\alpha}^{(t)}, \boldsymbol{\bar{S}}^{(t)}\right) & = -\ln  \Pr \left(\boldsymbol{y}_{\mathrm{T}} | \boldsymbol{\alpha}^{(t)}, \boldsymbol{\bar{S}}^{(t)}; \gamma^{(t)}\right) \\ \label{29}
	& = \ln |\boldsymbol{C}| + \operatorname{tr}\left\{\boldsymbol{C}^{-1} {\boldsymbol{R}}_y\right\} + \text{Const}.
\end{align}

\begin{proposition}
	Based on \eqref{29}, the logarithmic objective function ${\mathcal{L}} \left( \boldsymbol{\alpha}^{(t)}, \boldsymbol{\bar{S}}^{(t)}\right)$ can be further decoupled into the influence of $p$-th grid component $\mathcal{L} \left( \alpha_p^{(t)}, \bar{S}_p^{(t)} \right)$ and the influence of other grid components without $p$-th gird component $\mathcal{L} \left( \boldsymbol{\alpha}_{-p}^{(t)}, \boldsymbol{\bar{S}}_{-p}^{(t)} \right) $, i.e.,
	\begin{align}\label{30}
		{\mathcal{L}} \left( \boldsymbol{\alpha}^{(t)}, \boldsymbol{\bar{S}}^{(t)}\right) = \mathcal{L} \left( \boldsymbol{\alpha}_{-p}^{(t)}, \boldsymbol{\bar{S}}_{-p}^{(t)} \right)  + \mathcal{L} \left( \alpha_p^{(t)}, \bar{S}_p^{(t)} \right) + \text{Const},
	\end{align}
	where $\boldsymbol{\alpha}_{-p}^{(t)}$ denotes $\boldsymbol{\alpha}^{(t)}$ without the $p$-th element and $\boldsymbol{\bar{S}}_{-p}^{(t)}$  denotes $\boldsymbol{\bar{S}}^{(t)}$ without the $p$-th virtual sampling grid,
	$\mathcal{L} \left( \boldsymbol{\alpha}_{-p}^{(t)}, \boldsymbol{\bar{S}}_{-p}^{(t)} \right)$ is given by
	\begin{align}\label{31}
		\mathcal{L} \left( \boldsymbol{\alpha}_{-p}^{(t)}, \boldsymbol{\bar{S}}_{-p}^{(t)} \right) = \ln \left|\boldsymbol{C}_{-p}\right| + \operatorname{tr}\left\{\boldsymbol{C}_{-p}^{-1} {\boldsymbol{R}}_y\right\}
	\end{align}
	with 
	\begin{align}\nonumber
		\boldsymbol{C}_{-p} & = \boldsymbol{C} - \alpha_p^{(t)} \boldsymbol{\phi} (\bar{S}_p^{(t)}) \boldsymbol{\phi}^{\text{H}} (\bar{S}_p^{(t)}) \\ \label{32}
		& =  {\boldsymbol{\Phi}_{-p}} (\boldsymbol{\bar{S}}_{-p}^{(t)}) \boldsymbol{\Lambda}_{-p}^{(t)} {\boldsymbol{\Phi}_{-p}^{\text{H}}} (\boldsymbol{\bar{S}}_{-p}^{(t)}) + \left( \gamma^{(t)} \right)^{-1}  \boldsymbol{I}_{M_{\text{T}}},
	\end{align}
	${\boldsymbol{\Phi}_{-p}} (\boldsymbol{\bar{S}}_{-p}^{(t)})$ denotes the matrix $\boldsymbol{\Phi} (\boldsymbol{\bar{S}}^{(t)})$ without the $p$-th column, $\boldsymbol{\Lambda}_{-p}^{(t)}$ denotes the matrix $\boldsymbol{\Lambda}^{(t)}$ without the $p$-th row and $p$-th column, and $\mathcal{L} \left( \alpha_p^{(t)}, \bar{S}_p^{(t)} \right)$ is given by
	\begin{align}\label{33}
		\mathcal{L} \left( \alpha_p^{(t)}, \bar{S}_p^{(t)} \right) \! = \! \ln \left( 1 + \alpha_p^{(t)} Z (\bar{S}_p^{(t)}) \right) - \frac{Q (\bar{S}_p^{(t)})}{ 1 /  \alpha_p^{(t)}  + Z (\bar{S}_p^{(t)})}
	\end{align}
	with 
	\begin{subequations}\label{34}
		\begin{align}
			Z (\bar{S}_p^{(t)}) & =   \boldsymbol{\phi}^{\text{H}}(\bar{S}_p^{(t)}) \boldsymbol{C}_{-p}^{-1} \boldsymbol{\phi} (\bar{S}_p^{(t)}), \\
			Q (\bar{S}_p^{(t)}) & = \boldsymbol{\phi}^{\text{H}}(\bar{S}_p^{(t)}) \boldsymbol{C}_{-p}^{-1} {\boldsymbol{R}}_y \boldsymbol{C}_{-p}^{-1} \boldsymbol{\phi} (\bar{S}_p^{(t)}).
		\end{align}
	\end{subequations}
\end{proposition}
\emph{Proof:} See Appendix A. \hfill $\blacksquare$

To estimate the off-grid Doppler component around the $p$-th grid, we regard the hyper-parameter $ {\alpha}_p^{(t)}$ and virtual sampling grid point $\bar{S}_p^{(t)}$ as variables. For clarity, we denote the variables $ {\alpha}_p^{(t)}$ and $\bar{S}_p^{(t)}$ as $ {\alpha}_p $ and $\bar{S}_p $ with constraint ${\alpha}_p \geq 0$ and
$\bar{S}_p \in \mathcal{A}_p^{(t)} $, where $ \mathcal{A}_p^{(t)} = \left\lbrace \{\bar{\ell}_p^{(t)}, \bar{k}_p^{(t)} + \bar{\beta}_p\} \left| \bar{\beta}_p \in [-\frac{1}{2} r_{\nu}, \frac{1}{2} r_{\nu}] \right.  \right\rbrace $ is the local refinement grid around the $p$-th grid point. Then, the virtual sampling grid point $\bar{S}_p^{(t+1)}$ can be updated by minimizing  the object function $\mathcal{L}({\alpha}_p,  \bar{S}_p)$, i.e.,
\begin{align}\nonumber
	&   \bar{S}_p^{(t+1)}  = \mathop{\arg\min}_{{\alpha}_p \geq 0, \bar{S}_p \in \mathcal{A}_p^{(t)}}  \mathcal{L}({\alpha}_p,  {\bar{S}}_{p})  \label{35} \\
	&  = \mathop{\arg\min}_{{\alpha}_p \geq 0, \bar{S}_p \in \mathcal{A}_p^{(t)}} \ln \left( 1 +\ \alpha_p Z (\bar{S}_p) \right) - \frac{Q (\bar{S}_p)}{ 1 / \alpha_p  + Z (\bar{S}_p)}.
\end{align}

By letting $\frac{\partial \mathcal{L}({\alpha}_p,  {\bar{S}}_{p}) }{\partial {\alpha}_p}  = 0$ with constraint ${\alpha}_p \geq 0$, the hyper-parameter $\hat{\alpha}_p$ can be optimized as a function of virtual grid $\bar{S}_{p}$, i.e.,
\begin{align}\label{36}
	\hat{\alpha}_p =
	\begin{cases}
		\frac{Q(\bar{S}_p) - Z(\bar{S}_p)}{Z^2 (\bar{S}_p)}, & Q(\bar{S}_p)> Z(\bar{S}_p)\\
		0, & Q(\bar{S}_p) \leq Z(\bar{S}_p)
	\end{cases}.
\end{align}

We then substitute the results of $\hat{\alpha}_p$ in \eqref{36} into the object function $\mathcal{L}(\hat{\alpha}_p, {\bar{S}}_{p})$, i.e.,
\begin{align}\nonumber
	\mathcal{L}&(\hat{\alpha}_p, {\bar{S}}_{p}) \\\label{37}
	& =  
	\begin{cases}
		\log\left( \frac{Q(\bar{S}_p)}{Z(\bar{S}_p)}  \right)  - \frac{Q(\bar{S}_p)}{Z(\bar{S}_p)}  + 1, & Q(\bar{S}_p) > Z(\bar{S}_p)\\
		0, & Q(\bar{S}_p) \leq Z(\bar{S}_p)
	\end{cases}.
\end{align}

It is worth noting that the $\mathcal{L} (\hat{\alpha}_p, {\bar{S}}_{p})$ is a monotonically decreasing function of $\frac{Q(\bar{S}_p)}{ Z(\bar{S}_p)} $. Therefore, the optimal grid point can be obtained by maximizing $\frac{Q(\bar{S}_p)}{ Z(\bar{S}_p)} $.  However, it is intractable to directly obtain the closed-form solution of $\bar{S}_p$ to maximize $\frac{Q(\bar{S}_p)}{ Z(\bar{S}_p)} $.
To tackle this issue, we 
can quantify the off-grid Doppler component $\bar{\beta}_p$ from the $p$-th local refinement grid set $\mathcal{A}_p^{(t)}$ with a fine step size $\Delta$ and form the local quantify refinement grid $\bar{\mathcal{A}}_p^{(t)} =  \left\lbrace (\bar{\ell}_p^{(t)}, \bar{k}_p^{(t)} \!+\! \bar{\beta}_p) \left| \bar{\beta}_p \in [-\frac{1}{2} r_{\nu}, -\frac{1}{2} r_{\nu} \!+\! \Delta, ..., \frac{1}{2} r_{\nu}] \right.  \right\rbrace$ with size $|\bar{\mathcal{A}}_p^{(t)}| = r_{\nu} / \Delta + 1$.
Therefore, the $p$-th grid point $\bar{S}_p^{(t+1)}$ and the corresponding off-grid component $\bar{\beta}_p^{(t+1)}$ can be updated as
\begin{align}\label{38}
\bar{S}_p^{(t+1)} = \left\lbrace \bar{\ell}_p^{(t)}, \bar{k}_p^{(t)} + \bar{\beta}_p^{(t+1)} \right\rbrace  = \mathop{\arg\max}_{\bar{S}_p \in \bar{\mathcal{A}}_p^{(t)} } \frac{Q(\bar{S}_p)}{Z(\bar{S}_p)}.
\end{align}

Then, the delay and Doppler components can be updated as $\bar{\ell}_p^{(t+1)} = \bar{\ell}_p^{(t)}$, $\bar{k}_p^{(t+1)} = \bar{k}_p^{(t)} + \bar{\beta}^{(t+1)}_p$. 
Note that the proposed GR-SBL estimator employs a sequential iterative process that enables SBL and grid update to be performed in each iteration for improving iterative efficiency. Furthermore, we develop a parallel grid update strategy that enables the parallel update of multiple promising grid points $p \in \mathcal{P}$ to reduce processing latency.

\begin{algorithm}[t!]\color{black}
	\caption{Proposed Channel Estimation with SBL}
	\textbf{Input:} $\boldsymbol{y}_{\text{T}}$, $\varepsilon$,  $\epsilon$, $\rho, c, d$, $n_{{iter}}$ \\
	\textbf{Initialization:} $\boldsymbol{\bar{S}}^{(1)}$, $\boldsymbol{\Lambda}^{(1)} = \boldsymbol{I}_{M_{\text{T}}}$, $\gamma^{(1)} = \frac{100 M_{\text{T}}}{|| \boldsymbol{y}_{\text{T}} ||^2}$,  $\bar{P} = \lfloor \frac{M_{\text{T}}}{\ln(M_S)} \rfloor$, $t = 1$ \\
	\textbf{Repeat}
	\begin{algorithmic}[1]
		\STATE Set the off-grid Doppler component $\boldsymbol{\bar{\beta}}^{(t)} = \boldsymbol{0}_{M_S \times 1}$ and  formulate measurement matrix $\boldsymbol{\Phi} (\boldsymbol{\bar{S}}^{(t)}) $ and $\boldsymbol{\Psi} (\boldsymbol{\bar{S}}^{(t)})$ based on virtual grid $\boldsymbol{\bar{S}}^{(t)} $ by \eqref{15} and \eqref{40};
		\STATE Update the conditional posteriori variance $\boldsymbol{\Sigma}^{(t)}$ and mean $\boldsymbol{\mu}^{(t)}$  by \eqref{22} and \eqref{23}, respectively;
		\STATE Update the hyper-parameters $\boldsymbol{{\alpha}}^{(t+1)}$ and ${\gamma}^{(t+1)}$ by \eqref{26} and \eqref{27}, respectively;
		\STATE Extract $\bar{P}$ maximum elements from $\boldsymbol{{\alpha}}^{(t+1)}$ with grid index $p \in \mathcal{P}$;
		\STATE \textbf{Choose Grid Refinement or Grid Evolution:}
		\STATE ~~~$\boldsymbol{-}$ \textbf{Grid Refinement:} 
		Update the off-grid Doppler components $\bar{\beta}_p^{(t+1)}$  by \eqref{38},  $p \in \mathcal{P}$;
		\STATE ~~~$\boldsymbol{-}$ \textbf{Grid Evolution:} 
		Update the off-grid Doppler component $\bar{\beta}_p^{(t+1)}$  by \eqref{47}, $p \in \mathcal{P}$;
		\STATE Update the delay components, Doppler components and virtual sampling grid as $\bar{\ell}_p^{(t+1)} = \bar{\ell}_p^{(t)}$, $\bar{k}_p^{(t+1)} = \bar{k}_p^{(t)} + \bar{\beta}^{(t+1)}_p$ and $\bar{S}_p^{(t+1)} = \{\bar{\ell}_p^{(t+1)}, \bar{k}_p^{(t+1)}\} $,
		$p \in \mathcal{P}$; 
		\STATE  $ t = t + 1 $;
	\end{algorithmic}
	\textbf{Until} : $\frac{\left|\left|\boldsymbol{\alpha}^{(t+1)} - \boldsymbol{\alpha}^{(t)} \right|\right|_2 }{\left|\left| \boldsymbol{\alpha}^{(t)} \right|\right|_2} < \epsilon$ or $t = n_{{iter}}$; \\
	\\
	\textbf{Output} : Extract the index $\hat{p} \in \hat{\mathcal{P}}$ satisfying ${\alpha}_{\hat{p}}^{(t+1)} > \varepsilon$ with size $\hat{P}$; Output $\hat{P}$, $ \boldsymbol{\hat{h}} = \boldsymbol{\mu}^{(t)}_{\hat{\mathcal{P}}}$, $\boldsymbol{\hat{\ell}} = \boldsymbol{\bar{\ell}}_{\hat{\mathcal{P}}}^{(t+1)}$ and $\boldsymbol{\hat{k}} + \boldsymbol{\hat{\beta}} = \boldsymbol{\bar{k}}_{\hat{\mathcal{P}}}^{(t+1)}$.
	\label{alg1}
\end{algorithm} 

\subsection{GE-SBL for Off-Grid Doppler Component Estimation}

\begin{figure}[t!]
	\centering
	\includegraphics[scale=0.55]{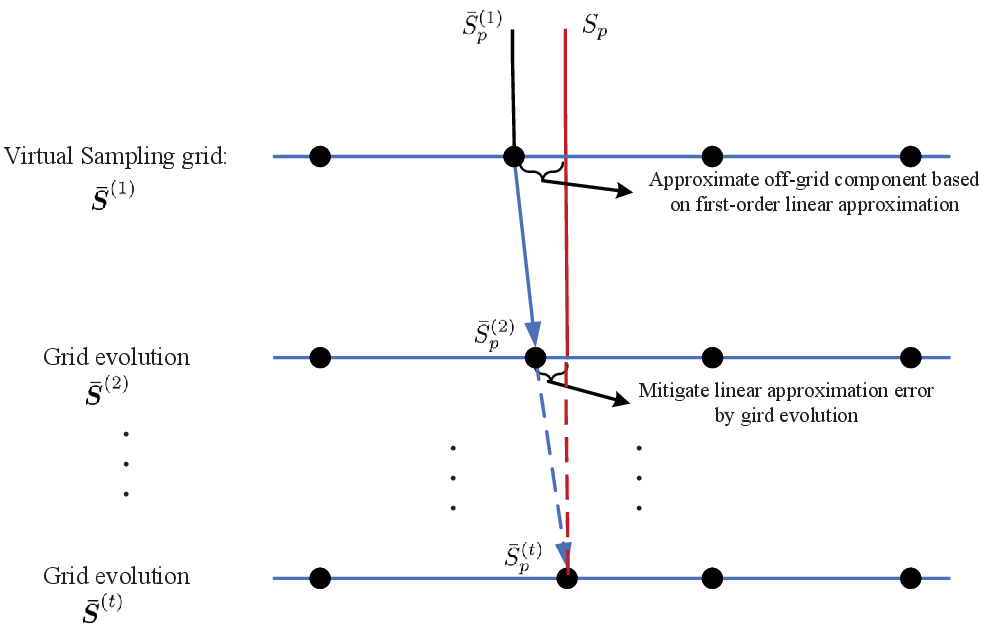}
	\caption{GE-SBL for off-grid Doppler component estimation.}
	\label{Fig_3}
\end{figure} 

Although the GR-SBL can effectively refine the local grid to mitigate off-grid component estimation errors, the estimation accuracy strongly depends on the search step sizes, and the fine search step sizes will result in a significant search burden. 
To overcome this limitation, we develop a grid evolution method in this sub-section and propose the GE-SBL estimator. 
By approximating the off-grid Doppler components based on first-order linear approximation and gradually adjusting the virtual sampling grid, the off-grid channel taps can finally match to the updated virtual grid. In addition, the proposed GE-SBL estimator can also mitigate the linear approximation error by gradually grid evolution and overcome the performance limitation of OG-SBL estimator \cite{b25}, which effectively enhances channel estimation accuracy. Note that our proposed GE-SBL estimator aims to gradually adjust the virtual sampling grid without adding new grid point, which is fundamentally differ from the grid evolution in \cite{b27.1} that performs grid fission for improving local grid resolution.
The grid evolution process is described in Fig. \ref{Fig_3}.

Specifically, the off-grid Doppler components are first approximated based on a first-order linear approximation, such that the measurement matrix can be re-expressed as
\begin{align}\label{39}
	\boldsymbol{\tilde{\Phi}} (\boldsymbol{\bar{S}})  = \boldsymbol{\Phi} (\boldsymbol{\bar{S}})  + \boldsymbol{\Psi} (\boldsymbol{\bar{S}})   \text{diag} \{\boldsymbol{\bar{\beta}}\},
\end{align}
where 
\begin{align}\label{40}
	\boldsymbol{\Psi} (\boldsymbol{\bar{S}}) = \left[ \boldsymbol{\psi} (\bar{S}_0), \boldsymbol{\psi} (\bar{S}_1)  ..., \boldsymbol{\psi} (\bar{S}_i), ..., \boldsymbol{\psi} (\bar{S}_{M_S - 1}) \right] 
\end{align}
with $ \boldsymbol{\psi} \left( \bar{S}_i \right) = \partial \boldsymbol{\phi} \left( \bar{S}_i \right) / \partial \bar{k}_i  $. According to the SBL principle, we obtain the conditional posterior distribution $ \Pr (\boldsymbol{\bar{h}} |\boldsymbol{y}_{\text{T}}; \boldsymbol{\alpha}^{(t)}, \gamma^{(t)}, \boldsymbol{S}^{(t)} ) $ with variance $\boldsymbol{\Sigma}^{(t)}$ and mean  $\boldsymbol{\mu}^{(t)}$ by \eqref{22} and \eqref{23}.\footnote{\label{n2}Note that the off-grid Doppler components are set $\boldsymbol{\bar{\beta}}^{(t)} = \boldsymbol{0}_{M_S \times 1}$ and we have $\boldsymbol{\tilde{\Phi}} (\boldsymbol{\bar{S}}^{(t)}) = 	\boldsymbol{\Phi} (\boldsymbol{\bar{S}}^{(t)})$  at the beginning of the $(t)$-th iteration. Then, the conditional posterior variance $\boldsymbol{\Sigma}^{(t)}$ and mean $\boldsymbol{\mu}^{(t)}$ based on $\boldsymbol{\tilde{\Phi}} (\boldsymbol{\bar{S}}^{(t)})$ are the same as \eqref{22} and \eqref{23} based on $\boldsymbol{\Phi} (\boldsymbol{\bar{S}}^{(t)})$.} The corresponding hyper-parameters $\boldsymbol{{\alpha}}^{(t+1)}$ and ${\gamma}^{(t+1)}$ can be obtained by \eqref{26} and \eqref{27} based on the EM algorithm. 

\begin{proposition}
Based on \eqref{39}, the optimization of off-grid Doppler component is equivalent to  
\begin{align}\label{41}
	& \boldsymbol{\bar{\beta}}^{(t+1)} \! = \! \arg\min_{\boldsymbol{\bar{\beta}}}  \mathbb{E}_{\Pr (\boldsymbol{\bar{h}} |\boldsymbol{y}_{\text{T}}; \boldsymbol{\alpha}^{(t)}, \gamma^{(t)}, \boldsymbol{\bar{S}}^{(t)} )} \!\left\lbrace \left| \left| \boldsymbol{y}_{\text{T}} - \boldsymbol{\tilde{\Phi}} (\boldsymbol{\bar{S}}^{(t)}) \boldsymbol{\bar{h}} \right|\right|^2 \! \right\rbrace  \\ \label{42} 
	& = \arg\min_{\boldsymbol{\bar{\beta}}} \left( \boldsymbol{\bar{\beta}}^{\text{T}} \boldsymbol{A}^{(t)} \boldsymbol{\bar{\beta}} - 2 \left( \boldsymbol{b}^{(t)} \right)^{\text{T}}  \boldsymbol{\bar{\beta}} \right) + \text{Const},
\end{align}
where $\boldsymbol{A}^{(t)} $ and $\boldsymbol{b}^{(t)}$ are given by \eqref{43} and \eqref{44}, shown at the bottom of the next page. 
\end{proposition}
\emph{Proof:} See Appendix B. \hfill $\blacksquare$
\begin{figure*}[hb]
	\hrulefill
	\begin{align}\label{43}
		\boldsymbol{A}^{(t)} & = \mathcal{R} \left\lbrace \left( \boldsymbol{\Psi}^{\text{H}} (\boldsymbol{\bar{S}}^{(t)}) \boldsymbol{\Psi} (\boldsymbol{\bar{S}}^{(t)})\right)^{*}  \odot \left( \boldsymbol{\mu}^{(t)}  \left( \boldsymbol{\mu}^{(t)} \right)^{\text{H}} + \boldsymbol{\Sigma}^{(t)} \right)  \right\rbrace. \\ \label{44}
		\boldsymbol{b}^{(t)} & = \mathcal{R} \left\lbrace \text{diag} \left\lbrace \boldsymbol{\mu}^{(t)}\right\rbrace  \boldsymbol{\Psi}^{\text{H}} (\boldsymbol{\bar{S}}^{(t)}) \left( \boldsymbol{y}_{\text{T}} - \boldsymbol{\Phi} (\boldsymbol{\bar{S}}^{(t)}) \boldsymbol{\mu}^{(t)} \right)  - \text{diag} \left\lbrace \boldsymbol{\Psi}^{\text{H}} (\boldsymbol{\bar{S}}^{(t)}) \boldsymbol{\Phi}(\boldsymbol{\bar{S}}^{(t)}) \boldsymbol{\Sigma}^{(t)} \right\rbrace \right\rbrace. 
	\end{align}
\end{figure*}

Since the off-grid Doppler component  $\boldsymbol{\bar{\beta}}$ and channel vector $\boldsymbol{\bar{h}}$ have the same non-zeros element positions, we can only estimate the $\bar{P}$ off-grid components with index $\mathcal{P}$ for reducing complexity, i.e.,  $\boldsymbol{\bar{\beta}}_{\text{T}} = \boldsymbol{\bar{\beta}}_{\mathcal{P}}  \in \mathbb{C}^{\bar{P} \times 1}$. Then, the corresponding $\boldsymbol{A}_{\text{T}}^{(t)}$ can be obtained by extracting $\mathcal{P}$ row and $\mathcal{P}$ column of $\boldsymbol{A}^{(t)}$, i.e., $\boldsymbol{A}_{\text{T}}^{(t)} = \boldsymbol{A}_{\mathcal{P}, \mathcal{P}}^{(t)}  \in \mathbb{C}^{\bar{P} \times \bar{P}} $, $\boldsymbol{b}_{\text{T}}^{(t)}$ can be obtained by extracting $\mathcal{P}$ elements of $\boldsymbol{b}^{(t)}$, i.e., $ \boldsymbol{b}_{\text{T}}^{(t)} = \boldsymbol{b}_{\mathcal{P}}^{(t)}  \in \mathbb{C}^{\bar{P} \times 1}$. When $\boldsymbol{A}_{\text{T}}^{(t)}$ is invertible, the off-grid Doppler components $\boldsymbol{\bar{\beta}}_{\text{T}}^{(t+1)}$ can be obtained based on \eqref{42} as
\begin{align}\label{45}
	\boldsymbol{\bar{\beta}}_{\text{T}}^{(t+1)} = \left(  \boldsymbol{A}_{\text{T}}^{(t)} \right)^{-1}  \boldsymbol{b}_{\text{T}}^{(t)}.
\end{align}

Otherwise, we can perform the element-wise update for $\boldsymbol{\bar{\beta}}_{\text{T}}^{(t+1)}$, i.e.,
\begin{align}\label{46}
	{\bar{\beta}}_{\text{T}, j}^{(t+1)}  = \frac{ {b}^{(t)}_{\text{T}, j} - \left( \boldsymbol{A}_{\text{T}, \{j, -j \}}^{(t)} \right)^{\text{T}}  \boldsymbol{\bar{\beta}}^{(t)}_{\text{T}, -j} }{ {A}_{\text{T},\{j,j\}}^{(t)}  },
\end{align}
where $\boldsymbol{A}_{\text{T}, \{j, -j\}}^{(t)} $ denotes the $j$-th row of matrix  $\boldsymbol{A}_{\text{T}}^{(t)}$, excluding the $j$-th column.
$\boldsymbol{\bar{\beta}}^{(t)}_{\text{T}, -j}$ denotes the $\boldsymbol{\bar{\beta}}^{(t)}_{\text{T}}$ without the $j$-th element, $j = 1, 2, ..., \bar{P}$.

According to the priori constraint of off-grid Doppler components in \eqref{20},  $\boldsymbol{\bar{\beta}}_{\text{T}}^{(t+1)}$ can be finally updated as
\begin{align}\label{47}
	{\bar{\beta}}_{\text{T},j}^{(t+1)}  = 
	\begin{cases}
		{\bar{\beta}}_{\text{T}, j}^{(t+1)} , & {\bar{\beta}}_{\text{T}, j}^{(t+1)}  \in \left[ -\frac{1}{2} r_{\nu}, \frac{1}{2} r_{\nu}  \right] \\
		-\frac{1}{2} r_{\nu}, & 	{\bar{\beta}}_{\text{T}, j}^{(t+1)}  < -\frac{1}{2} r_{\nu} \\
		\frac{1}{2} r_{\nu}, & \text{otherwise}
	\end{cases}.
\end{align}

Then, we can recover the off-grid Doppler components $\boldsymbol{\bar{\beta}}_{\mathcal{P}}^{(t+1)}$ based on  $\boldsymbol{\bar{\beta}}_{\text{T}}^{(t+1)}$ and the truncation index set $\mathcal{P}$, i.e., $\boldsymbol{\bar{\beta}}_{\mathcal{P}}^{(t+1)} = \boldsymbol{\bar{\beta}}_{\text{T}}^{(t+1)}$. The 
delay and Doppler components can be updated as $\bar{\ell}_p^{(t+1)} = \bar{\ell}_p^{(t)}$ and $\bar{k}_p^{(t+1)} = \bar{k}_p^{(t)} + \bar{\beta}^{(t+1)}_p$, and the corresponding virtual sampling grid can be updated as $\bar{S}_p^{(t+1)} = \{\bar{\ell}_p^{(t+1)}, \bar{k}_p^{(t+1)}\}  $, $p \in \mathcal{P}$.

The GR-SBL and GE-SBL iterative algorithms terminate when $  \frac{\left|\left|\boldsymbol{\alpha}^{(t+1)} - \boldsymbol{\alpha}^{(t)} \right|\right|_2 }{\left|\left| \boldsymbol{\alpha}^{(t)} \right|\right|_2} < \epsilon$ with a small value $\epsilon$ or the maximum iteration number $n_{iter}$ is reached. We then extract the index $\hat{p} \in \hat{\mathcal{P}}$ satisfying ${\alpha}_{\hat{p}}^{(t+1)} > \varepsilon$, where $\hat{\mathcal{P}}$ is the index set of $\hat{p}$ with size  $\hat{P}$, $\varepsilon$ is the threshold. Finally, the number of paths can be estimated as $\hat{P}$. The channel gain can be estimated as  $ \boldsymbol{\hat{h}} = \boldsymbol{\mu}^{(t)}_{\hat{\mathcal{P}}}$, and the corresponding delay and Doppler can be estimated as $\boldsymbol{\hat{\ell}} = \boldsymbol{\bar{\ell}}_{\hat{\mathcal{P}}}^{(t+1)}$ and $\boldsymbol{\hat{k}} + \boldsymbol{\hat{\beta}} = \boldsymbol{\bar{k}}_{\hat{\mathcal{P}}}^{(t+1)}$. Based on the estimated channel parameters, the channel matrix  $\boldsymbol{\hat{H}}$ can be reconstructed by \eqref{8}.
The proposed channel estimation with SBL can be summarized in \textbf{Algorithm 1}.

\vspace{-8pt}
\subsection{Genie Bound}

From the above discussions, it can be observed that the performance of the proposed GR-SBL and GE-SBL estimators mainly depend on the accuracy of dynamic grid update. To further evaluate the effectiveness of the proposed off-grid channel estimation methods in the AFDM system, the genie bound is usually employed as a performance baseline of the off-grid sparse signal recovery in \cite{b25, b33.1, b33.2}. It assumes that the perfect dynamic grid updates are performed, i.e., the delay and the Doppler indices are on the predefined virtual sampling grid. In other words, there is no off-grid component to be estimated, thereby simplifying the analysis and providing a performance baseline for off-grid sparse signal recovery.

Specifically, we can formulate the measurement matrix $\boldsymbol{\Phi}(\boldsymbol{\bar{S}})$ based on the perfectly delay and Doppler sampling grid $\boldsymbol{\bar{S}}$. Then, the conditional posterior variance $\boldsymbol{\Sigma}^{(t)}$ and mean $\boldsymbol{\mu}^{(t)}$ can be updated by \eqref{22} and \eqref{23}. The corresponding hyper-parameters $\boldsymbol{\alpha}^{(t+1)}$ and noise precision $\gamma^{(t+1)}$ can be updated by \eqref{26} and \eqref{27}.

%\textcolor{blue}{To evaluate the effectiveness of the proposed off-grid channel estimation methods in the AFDM system,  the Cramer-Rao Lower Bound (CRLB) serves as a fundamental theoretical benchmark for the mean squared error of channel coefficient estimates. However, deriving the CRLB for off-grid channel estimation is particularly challenging due to the coupling of unknown off-grid Doppler components in the measurement matrix. As a result, obtaining the likelihood function $ \Pr (\boldsymbol{y}_{\text{T}} |\boldsymbol{\bar{h}}, \boldsymbol{\beta}) $ and the  corresponding fisher information matrix (FIM) becomes highly complex.}

%\textcolor{blue}{As an alternative, the genie bound is usually employed to evaluate the performance of sparse signal recovery in \cite{b25, b33.1, b33.2}. It assumes that the perfect dynamic grid updates are performed, i.e., the delay and the Doppler indices are on the predefined virtual sampling grid. In other words, there is no off-grid component to be estimated, thereby simplifying the analysis and providing a baseline performance for sparse signal recovery. Specifically, we can formulate the measurement matrix $\boldsymbol{\Phi}(\boldsymbol{\bar{S}})$ based on the perfectly delay and Doppler sampling grid $\boldsymbol{\bar{S}}$. Then, the conditional posterior mean $\boldsymbol{\mu}^{(t)}$ and variance $\boldsymbol{\Sigma}^{(t)}$ can be updated by \eqref{22} and \eqref{23}. The corresponding hyper-parameters $\boldsymbol{\alpha}^{(t+1)}$ and noise precision $\gamma^{(t+1)}$ can be updated by \eqref{26} and \eqref{27}.}

\section{Proposed Low-Complexity Distributed Channel Estimation Scheme}
From \textbf{Algorithm 1}, we can observe that the proposed  GR-SBL and GE-SBL estimators consist of several core modules. Therefore, we can analyze the complexity of each module to determine the overall complexity in each iteration.

In the measurement matrix update, the complexity can be attributed to the computing of $\boldsymbol{\phi} (\bar{S}_p) $ for $\bar{P}$ grid in \eqref{13} and \eqref{15}, which is given by $(3 \bar{P} |\mathcal{M}_p| M_{\text{T}})$. 
For the update of posteriori variance $\boldsymbol{\Sigma}$ and mean $\boldsymbol{\mu}$, the complexity 
can be attributed to \eqref{22} and \eqref{23}, which are given by  $(M_{\text{T}}^3 + M_{\text{T}} M_{S} (2M_{\text{T}} + M_{S} +  2)) $ and $ (M_S^2 + M_{\text{T}} M_S) $. For the update of the hyper-parameters $\boldsymbol{\alpha}$ and $\gamma$, the 
complexity can be attributed to  \eqref{26} and \eqref{27}, which are given by $(3 M_S) $ and  $ (M_{\text{T}} M_S + M_{\text{T}} + M_S) $. For the update of off-grid Doppler component, we develop the grid refinement and grid evolution methods, respectively. In the grid refinement method, the complexity can be attributed to \eqref{34} and \eqref{38}, which are given by $ (\bar{P}|\bar{\mathcal{A}}_p| (4M_{\text{T}}^2 + 2 M_{\text{T}}))$ and $(\bar{P}|\bar{\mathcal{A}}_p|)$. In the grid evolution method, the complexity can be attributed to \eqref{45}, which is given by $(\bar{P}^3 + \bar{P}^2)$. Therefore, the overall complexity of GR-SBL and GE-SBL estimators can be obtained by the superposition of each module complexity, which are summarized in \textbf{Table \ref{tab2}}. 

\begin{table*}[t!]
	\renewcommand{\arraystretch}{1.5}
	\begin{center}
		\caption{Computational Complexity of Different Channel Estimation Schemes}
		\label{tab2}
		\begin{tabular}{c|c}
			\hline 
			\text{Algorithm} & \text{Computational complexity for each iteration} \\ 
			\hline 
			OG-SBL \cite{b25} & $ M_{\text{T}}^3 +  ( 2M_{\text{T}} + M_S + 3) M_{\text{T}} M_{S} + M_S^2 +  (M_S + 1)M_{\text{T}}  + 4 M_S + \bar{P}^3 + \bar{P}^2 $  \\
			GR-SBL & $ M_{\text{T}}^3 +  ( 2M_{\text{T}} + M_S + 3) M_{\text{T}} M_{S} + M_S^2 +  (3 \bar{P} |\mathcal{M}_p| + M_S + 1)M_{\text{T}}  + 4 M_S + \bar{P} |\bar{\mathcal{A}}_p| (4 M_{\text{T}}^2 + 2 M_{\text{T}} + 1) $ \\
			GE-SBL & $ M_{\text{T}}^3 +  ( 2M_{\text{T}} + M_S + 3) M_{\text{T}} M_{S} + M_S^2 +  (3 \bar{P} |\mathcal{M}_p| + M_S + 1)M_{\text{T}}  + 4 M_S + \bar{P}^3 + \bar{P}^2 $   \\
			D-GR-SBL  & $C(M_{\text{c}}^3 +  (2M_{\text{c}} + |\mathcal{J}_c| +  3)M_{\text{c}}  |\mathcal{J}_c| + |\mathcal{J}_c|^2) + (3 \bar{P} |\mathcal{M}_p| + M_S + 1 )M_{\text{T}}   + 4M_S +  \bar{P} |\bar{\mathcal{A}}_p| (4 M_{\text{T}}^2 + 2 M_{\text{T}} + 1) $ \\
			D-GE-SBL  & $C(M_{\text{c}}^3 +  (2M_{\text{c}} + |\mathcal{J}_c| +  3)M_{\text{c}}  |\mathcal{J}_c| + |\mathcal{J}_c|^2) + (3 \bar{P} |\mathcal{M}_p| + M_S + 1)M_{\text{T}}  + 4 M_S + \bar{P}^3 + \bar{P}^2  $  \\
			\hline
		\end{tabular}
	\end{center}
	\vspace{-10pt}
\end{table*}

From the above analysis, we can observe that the proposed GR-SBL and GE-SBL estimators mainly depend on the SBL framework and suffer from the high complexity caused by matrix inversion in the update of posteriori mean and variance.
Thus, how to reduce the complexity of SBL is important and challenging. To address this limitation, we develop a distributed computing scheme to decompose
the large dimensional channel estimation model into multiple manageable sub-models for complexity reduction, then yielding the corresponding D-GR-SBL and D-GE-SBL estimators, respectively. 

Specifically, we firstly decouple the channel estimation model \eqref{14} into $C$ group. Then, the received signal $\boldsymbol{y}_{\text{T}}$, measurement matrix $\boldsymbol{\Phi} (\boldsymbol{\bar{S}})$ and measurement noise $\boldsymbol{\omega}_{\text{T}}$ can be partitioned as
\begin{subequations}\label{48}
	\begin{align}
		\boldsymbol{y}_{\text{T}} & = [\boldsymbol{y}_{\text{T},1}^{\text{T}}, \boldsymbol{y}_{\text{T},2}^{\text{T}}, ..., \boldsymbol{y}_{\text{T},C}^{\text{T}}]^{\text{T}}, \\
		\boldsymbol{\Phi} (\boldsymbol{\bar{S}}) & = [\boldsymbol{\Phi}_{1}^{\text{T}} (\boldsymbol{\bar{S}}), \boldsymbol{\Phi}_{2}^{\text{T}} (\boldsymbol{\bar{S}}), ..., \boldsymbol{\Phi}_{C}^{\text{T}} (\boldsymbol{\bar{S}})]^{\text{T}} , \\
		\boldsymbol{\omega}_{\text{T}} & = [\boldsymbol{\omega}_{\text{T},1}^{\text{T}}, \boldsymbol{\omega}_{\text{T},2}^{\text{T}}, ..., \boldsymbol{\omega}_{\text{T},C}^{\text{T}}]^{\text{T}},
	\end{align}
\end{subequations}
where $\boldsymbol{y}_{\text{T},c} \in \mathbb{C}^{M_c \times 1}$, $\boldsymbol{\Phi}_{c} (\boldsymbol{\bar{S}}) \in \mathbb{C}^{M_c \times M_S} $ and $\boldsymbol{\omega}_{\text{T},c} \in \mathbb{C}^{M_c \times 1}$ with $M_c = M_{\text{T}} / C$, $c = 1, 2, ..., C$. The $c$-th group channel estimation model can be given by
\begin{align}\label{49}
	\boldsymbol{y}_{\text{T},c} = \boldsymbol{\Phi}_{c} (\boldsymbol{\bar{S}}) \boldsymbol{\bar{h}} + \boldsymbol{\omega}_{\text{T},c}. 
\end{align}

\begin{figure}[t!]
	\captionsetup{font = {scriptsize,scriptsize,scriptsize,small}, format = hang, justification = centering}
	\subfloat[ Doppler sampling resolution $r_{\nu} = 1$.]{\includegraphics[scale=0.3]{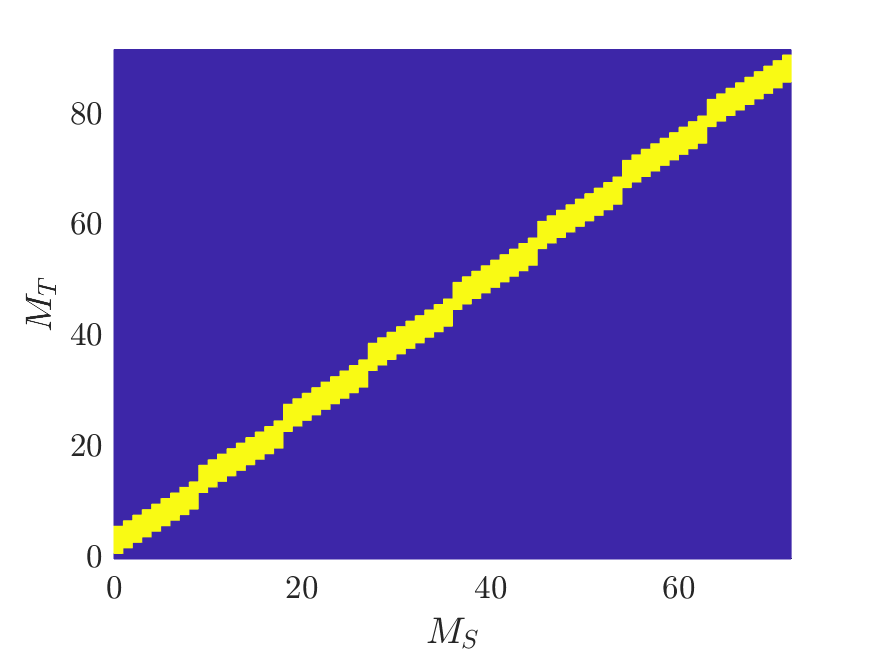}}
	%	\hfil
	\subfloat[Doppler sampling resolution $r_{\nu} = 0.5$.]{\includegraphics[scale=0.3]{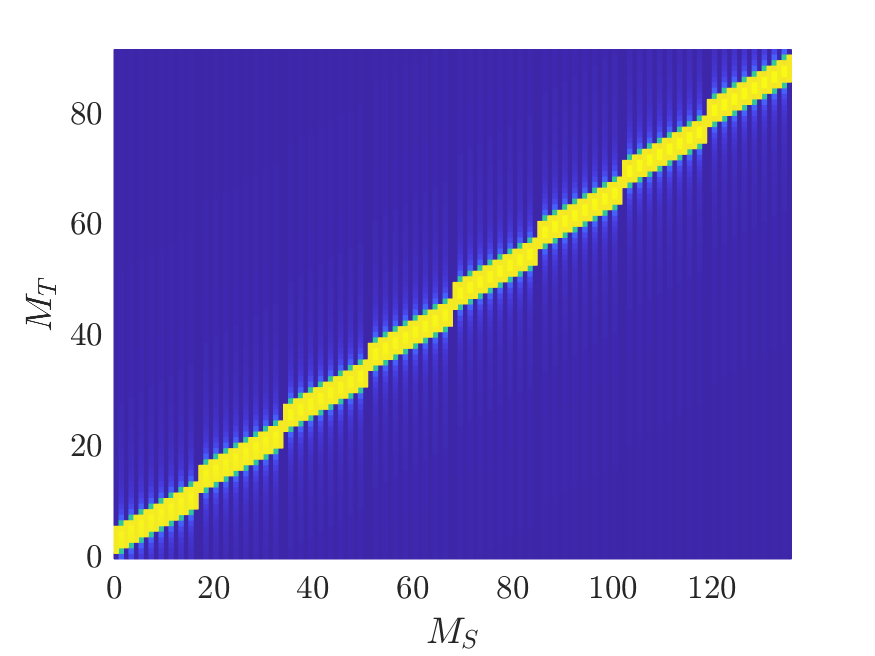}}
	\hfil
	\caption{AFDM measurement matrix structure $\boldsymbol{\Phi} (\boldsymbol{\bar{S}})$  with five pilot symbols.}
	\label{Fig_4}
\end{figure}

\begin{figure}[t!]
	\centering
	\includegraphics[scale=0.36]{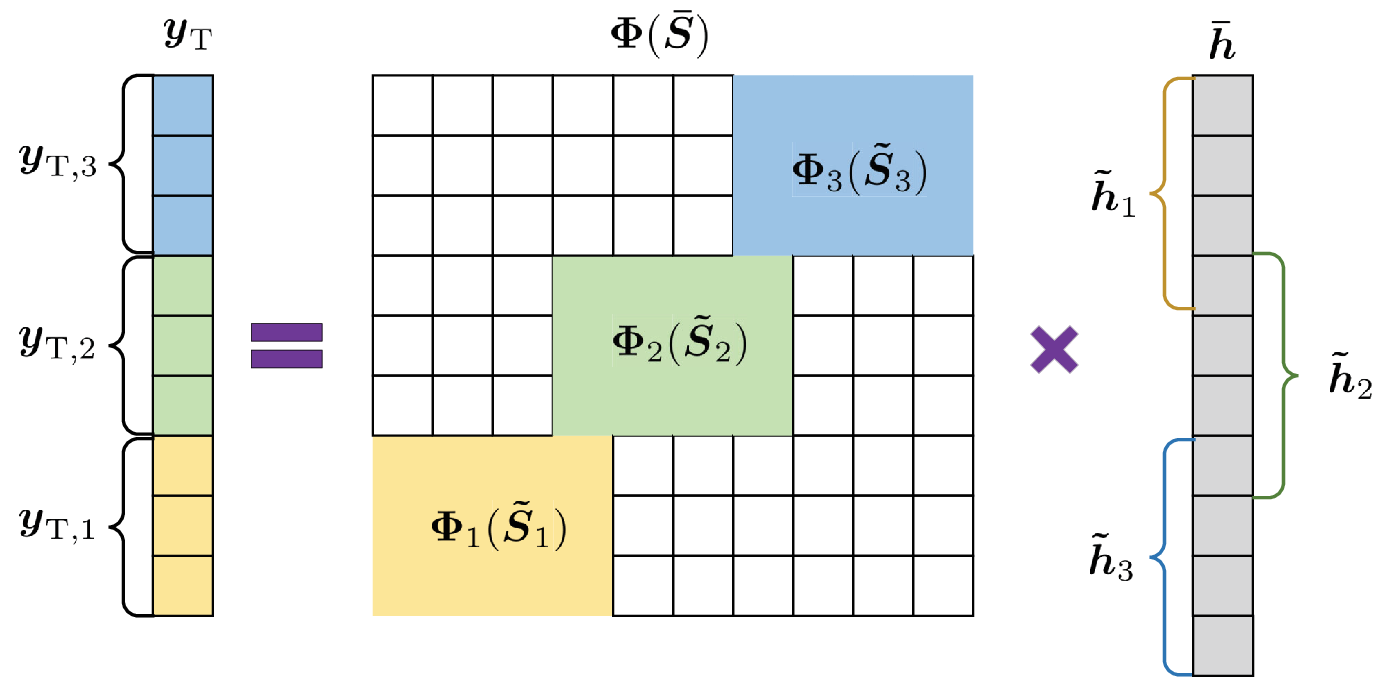}
	\caption{Distributed scheme with three groups as an example.}
	\label{Fig_5}
\end{figure}

According to \cite{b6}, the AFDM system can effectively resolve the paths with different delays based on the optimized chirp parameters, this characteristic accommodate a 
significantly sparse measurement matrix structure. To more clearly, Fig. \ref{Fig_4} provides the AFDM measurement matrix structure $\boldsymbol{\Phi} (\boldsymbol{\bar{S}}) \in \mathbb{C}^{M_{\text{T}} \times M_S}$ with different Doppler sampling resolutions $r_{\nu}$ established based on five pilot symbols, i.e., the AFDM measurement matrix structure $\boldsymbol{\Phi} (\boldsymbol{\bar{S}})$ with Doppler sampling resolutions $r_{\nu} = 1$ is depicted in Fig. \ref{Fig_4}(a) and $r_{\nu} = 0.5$ is depicted in Fig. \ref{Fig_4}(b).
It can be observed that the measurement matrix has a significant sparse structure.
Therefore, for the  
$c$-th group channel estimation model in \eqref{49}, only part of channel components in $\boldsymbol{\bar{h}}$ with index $\mathcal{J}_c$ are associated with $\boldsymbol{y}_{\text{T},c}$, i.e., $\boldsymbol{\tilde{h}}_c = \boldsymbol{\bar{h}}_{\mathcal{J}_c} \in \mathbb{C}^{|\mathcal{J}_c| \times 1}$, where $ |\mathcal{J}_c| $ is the size of $\mathcal{J}_c$. 
Then, we can further trim the $c$-th group measurement matrix $\boldsymbol{\Phi}_c (\boldsymbol{\bar{S}})$ by extracting corresponding $\mathcal{J}_c$ column into $\boldsymbol{{\Phi}}_c (\boldsymbol{\tilde{S}}_c) = \boldsymbol{\Phi}_c (\boldsymbol{\bar{S}}_{\mathcal{J}_c}) \in \mathbb{C}^{M_c \times |\mathcal{J}_c|} $.
Typically, the $|\mathcal{J}_c|$ is much smaller than $M_S$. Fig. \ref{Fig_5} intuitively illustrates the distributed channel estimation model with three groups as an example. 
The  $c$-th group channel estimation model can be reduced to
\begin{align}\label{50}
	\boldsymbol{y}_{\text{T},c} = \boldsymbol{{\Phi}}_c (\boldsymbol{\tilde{S}}_c) \boldsymbol{\tilde{h}}_c + \boldsymbol{\omega}_{\text{T},c}.
\end{align}

%\begin{figure}[t!]
%	\centering
%	\includegraphics[scale=0.5]{Measure.eps}
%	\caption{Measurement matrix structure for AFDM system with five pilot symbols pattern.}
%	\label{Fig_4}
%\end{figure} 

According to the SBL principle, the $c$-th group posteriori  variance $\boldsymbol{\tilde{\Sigma}}_c^{(t)}$ and mean $\boldsymbol{\tilde{\mu}}_c^{(t)}$ can be updated as 
\begin{align}\nonumber
	&\boldsymbol{\tilde{\Sigma}}^{(t)}_c  = \boldsymbol{\Lambda}^{(t)}_c  - \boldsymbol{\Lambda}^{(t)}_c  \boldsymbol{{\Phi}}^{\text{H}}_{c} (\boldsymbol{\tilde{S}}_c^{(t)}) \\ \label{51}
	& \times  \left( \!\left( \gamma^{(t)} \right)^{-1}  \!\!\!\boldsymbol{I}_{M_{c}} \!+\! \boldsymbol{{\Phi}}_{c} (\boldsymbol{\tilde{S}}_c^{(t)}) \boldsymbol{\Lambda}^{(t)}_c \boldsymbol{{\Phi}}^{\text{H}}_{c} (\boldsymbol{\tilde{S}}_c^{(t)})   \right)^{-1} \boldsymbol{{\Phi}}_{c} (\boldsymbol{\tilde{S}}_c^{(t)}) \boldsymbol{\Lambda}^{(t)}_c, \\ \label{52}
	& \boldsymbol{\tilde{\mu}}^{(t)}_c  = \gamma^{(t)} \boldsymbol{\tilde{\Sigma}}^{(t)}_c \boldsymbol{{\Phi}}^{\text{H}}_{c} (\boldsymbol{\tilde{S}}_c^{(t)}) \boldsymbol{y}_{\text{T},c}. 
\end{align}
where $\boldsymbol{\Lambda}_c^{(t)} = \text{diag} \{ \boldsymbol{\alpha}_c^{(t)} \}$ with $\boldsymbol{\alpha}_c^{(t)} =\boldsymbol{\alpha}^{(t)}_{\mathcal{J}_c}$. Then, we can combine the estimation results of multiple sub-models
based on Gaussian combining rule, where the mean $\boldsymbol{\mu}^{(t)}$ and variance $\boldsymbol{\Sigma}^{(t)}$ can be finally estimated as
\begin{align}\label{53}
	\Sigma_{\{i,i\}}^{(t)} & = \left( \sum_{\bar{c}_i \in {\bar{\mathcal{{C}}}}_i} \frac{1}{{\tilde{\Sigma}}_{\bar{c}_i, \{\bar{j}, \bar{j}\}}^{(t)} }\right)^{-1}, \\ \label{54}
	 \mu_i^{(t)} & = \Sigma_{\{i,i\}}^{(t)} \left( \sum_{\bar{c}_i \in \bar{\mathcal{C}}_i} \frac{\tilde{\mu}_{\bar{c}_i, \bar{j}}^{(t)} }{\tilde{\Sigma}_{\bar{c}_i, \{\bar{j}, \bar{j}\}}^{(t)} }\right), 
\end{align}
where $\bar{c}_i \in \mathcal{\bar{C}}_i$ is the index of $\bar{h}_i$  associated with $\boldsymbol{y}_{\text{T},\bar{c}_i}$, 
$\bar{j}$ is the corresponding position index in $\boldsymbol{y}_{\text{T},\bar{c}_i}$, and $\mathcal{\bar{C}}_i$ is the index set of $\bar{c}_i$.

Therefore, we can obtain the posteriori variance $\boldsymbol{\Sigma}^{(t)}$ and mean $\boldsymbol{\mu}^{(t)}$  by \eqref{53} and \eqref{54} based on distributed estimation. The hyper-parameters $\boldsymbol{\alpha}^{(t+1)}$ and noise precision $\gamma^{(t+1)}$ can be updated by \eqref{26} and \eqref{27} based on the posteriori variance $\boldsymbol{\Sigma}^{(t)}$ and mean $\boldsymbol{\mu}^{(t)}$. 
Then, the off-grid Doppler component of D-GR-SBL and D-GE-SBL estimators can be respectively estimated based on grid refinement method in \eqref{38} and grid evolution method in \eqref{47}, similar to GR-SBL and GE-SBL methods. 
The low-complexity distributed SBL channel estimation scheme is summarized in \textbf{Algorithm 2}.

\begin{algorithm}[t!]
	\caption{Proposed Low-Complexity Distributed SBL Channel Estimation Scheme}\label{alg2}
	\textbf{Input:} $\boldsymbol{y}_{\text{T}}$, $\varepsilon$,  $\epsilon$, $\rho, c, d$, $n_{{iter}}$ \\
	\textbf{Initialization:} $\boldsymbol{\bar{S}}^{(1)}$, $\boldsymbol{\Lambda}^{(1)} = \boldsymbol{I}_{M_{\text{T}}}$, $\gamma^{(1)} = \frac{100 M_{\text{T}}}{|| \boldsymbol{y}_{\text{T}} ||^2}$,  $\bar{P} = \lfloor \frac{M_{\text{T}}}{\ln(M_S)} \rfloor$, $t = 1$ \\
	\textbf{Repeat}
	\begin{algorithmic}[1]
		\STATE Set the off-grid Doppler component $\boldsymbol{\bar{\beta}}^{(t)} = \boldsymbol{0}_{M_S \times 1}$ and  formulate measurement matrix $\boldsymbol{\Phi} (\boldsymbol{\bar{S}}^{(t)}) $ and $\boldsymbol{\Psi} (\boldsymbol{\bar{S}}^{(t)})$ based on virtual grid $\boldsymbol{\bar{S}}^{(t)} $ by \eqref{15} and \eqref{40};
		\STATE Update $c$-th group conditional posteriori variance $\boldsymbol{\tilde{\Sigma}}^{(t)}_{c}$ and mean $\boldsymbol{\tilde{\mu}}^{(t)}_{c}$  by \eqref{51} and \eqref{52}, respectively, $c = 1, 2, ..., C$;
		\STATE Generate the posteriori variance $\boldsymbol{\Sigma}^{(t)}$ and mean $\boldsymbol{\mu}^{(t)}$ based on Gaussian combining rule by \eqref{53} and \eqref{54};
		\STATE Update the hyper-parameters $\boldsymbol{{\alpha}}^{(t+1)}$ and ${\gamma}^{(t+1)}$ by \eqref{26} and \eqref{27}, respectively;
		\STATE Extract $\bar{P}$ maximum elements from $\boldsymbol{{\alpha}}^{(t+1)}$ with grid index $p \in \mathcal{P}$;
		\STATE \textbf{Choose Grid Refinement or Grid Evolution:}
		\STATE ~~~$\boldsymbol{-}$ \textbf{Grid Refinement:} 
		Update the off-grid Doppler components $\bar{\beta}_p^{(t+1)}$ by \eqref{38},  $p \in \mathcal{P}$;
		\STATE ~~~$\boldsymbol{-}$ \textbf{Grid Evolution:} 
		Update the off-grid Doppler component $\bar{\beta}_p^{(t+1)}$  by \eqref{47}, $p \in \mathcal{P}$;
		\STATE Update the delay components, Doppler components and virtual sampling grid as $\bar{\ell}_p^{(t+1)} = \bar{\ell}_p^{(t)}$, $\bar{k}_p^{(t+1)} = \bar{k}_p^{(t)} + \bar{\beta}^{(t+1)}_p$ and $\bar{S}_p^{(t+1)} = \{\bar{\ell}_p^{(t+1)}, \bar{k}_p^{(t+1)}\} $,
		$p \in \mathcal{P}$; 
		\STATE  $ t = t + 1 $;
		\end{algorithmic}
		\textbf{Until} : $\frac{\left|\left|\boldsymbol{\alpha}^{(t+1)} - \boldsymbol{\alpha}^{(t)} \right|\right|_2 }{\left|\left| \boldsymbol{\alpha}^{(t)} \right|\right|_2} < \epsilon$ or $t = n_{{iter}}$; \\
		\\
		\textbf{Output} : Extract the index $\hat{p} \in \hat{\mathcal{P}}$ satisfying ${\alpha}_{\hat{p}}^{(t+1)} > \varepsilon$ with size $\hat{P}$; Output $\hat{P}$, $ \boldsymbol{\hat{h}} = \boldsymbol{\mu}^{(t)}_{\hat{\mathcal{P}}}$, $\boldsymbol{\hat{\ell}} = \boldsymbol{\bar{\ell}}_{\hat{\mathcal{P}}}^{(t+1)}$ and $\boldsymbol{\hat{k}} + \boldsymbol{\hat{\beta}} = \boldsymbol{\bar{k}}_{\hat{\mathcal{P}}}^{(t+1)}$.
		\label{alg2}
\end{algorithm} 

Based on distributed computing scheme, the complexity of posteriori variance $\boldsymbol{\Sigma}_c^{(t)}$ and mean $\boldsymbol{\mu}_c^{(t)}$  can be attributed to \eqref{51} and \eqref{52}, then the complexity of all groups is given by $ (C(M_{\text{c}}^3 + (2M_{\text{c}} + |\mathcal{J}_c| +  2)M_{\text{c}}  |\mathcal{J}_c| )) $ and $ (C ( |\mathcal{J}_c|^2 + M_{\text{c}} |\mathcal{J}_c| )) $.   Typically, $ M_{\text{c}}$ and $|\mathcal{J}_c|$ are much smaller than $ M_{\text{T}}$ and $M_{S}$, which can significantly reduce the computational complexity compared to original SBL framework. In addition, the 
parallel computing for each group is also supported, which can further reduce the computational latency. Finally, the overall complexity of the proposed D-GR-SBL and D-GE-SBL estimators is summarized in \textbf{Table \ref{tab2}}.

\section{Simulation Results}
In this section, we comprehensively evaluate the  channel estimation performance of our proposed GR-SBL, GE-SBL and corresponding distributed estimation schemes in AFDM systems over doubly-dispersive channels.
The normalized mean squared error (NMSE) is employed as the evaluation criterion for channel estimation performance, given by  $10\log_{10}\left( \frac{|\boldsymbol{H} - \boldsymbol{\hat{H}}|^2}{|\boldsymbol{H}|^2}\right) $. 
We consider the QPSK modulated AFDM systems. The carrier frequency is $f_c = 4$GHz with sub-carrier space $\varDelta f = 15$kHz. The normalized
maximum delay and Doppler are given by $l_{max} = 7$ and $k_{max} = 3$, respectively.
The normalized Doppler shift of each delay is generated based on Jakes’ formulation, i.e., $k_{i_p} = k_{max} \cos(\theta_{i_p})$, where $\theta_{i_p}$ is uniformly distributed over $[-\pi, \pi]$. The delay and Doppler sampling resolution are $r_{\tau} = 1$ and $r_{\nu} = 1$, respectively.
Without loss of generality, we consider 5 pilot symbols and set the total pilot power to be 30dB higher than that of the data symbols\footnote{\label{n4} Note that the pilot power is set 30dB higher than that of the data symbols is just an example to approximately show the performance bound of AFDM system, thereby better unlocking the performance potential of AFDM system in high-mobility scenarios, as done in \cite{b25, b26, b27, b27.1}. Furthermore, our proposed algorithms can be also extended to other pilot-to-data power ratio configurations for practical requirements.}.
The signal-to-noise ratio (SNR) is defined as the ratio of the average power of the data symbols to the noise power. For the parameters in algorithms, we set the root parameters $\rho = 10^{-2}$, $c = d = 10^{-6}$, the thresholds are $\varepsilon = 10^{-3}$ and $\epsilon = 10^{-4}$. The maximum number of iterations is $n_{iter} = 100$.

\begin{figure}[t]
	\centering
	\includegraphics[scale=0.55]{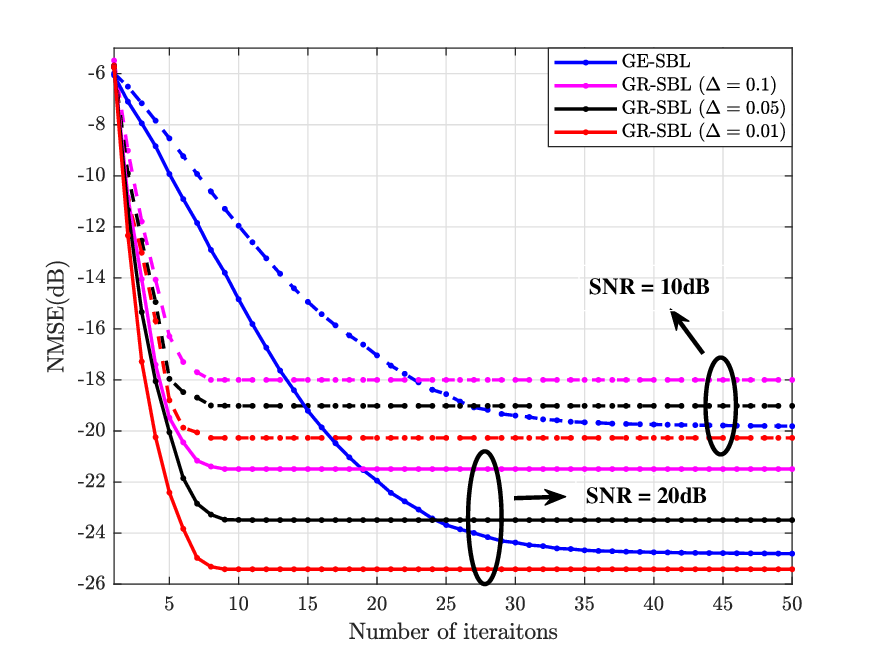}
	\caption{Convergence analysis versus different iteration numbers.}
	\label{Fig_6}
\end{figure}

We first analyze the convergence behavior of the proposed GR-SBL and GE-SBL estimators with different SNRs in Fig. \ref{Fig_6}. It is obvious that each estimator converges after a certain number of iterations and the convergence performance benefits from higher SNR. The GR-SBL estimator with a fine search step size such as $\Delta = 0.01$ shows impressive estimation performance, but the performance significantly degrades as the step size increases. However, the fine step size will result in a high search burden. We also notice that the performance of the  proposed GE-SBL estimator can approach that of GR-SBL with fine step size $\Delta = 0.01$, achieving a favorable balance in estimation accuracy and computational efficiency.

\begin{figure}[t!]
	\captionsetup{font = {scriptsize,scriptsize,scriptsize,small}, format = hang, justification = centering}
	\subfloat[Complexity comparison of different estimators.]{\includegraphics[scale=0.55]{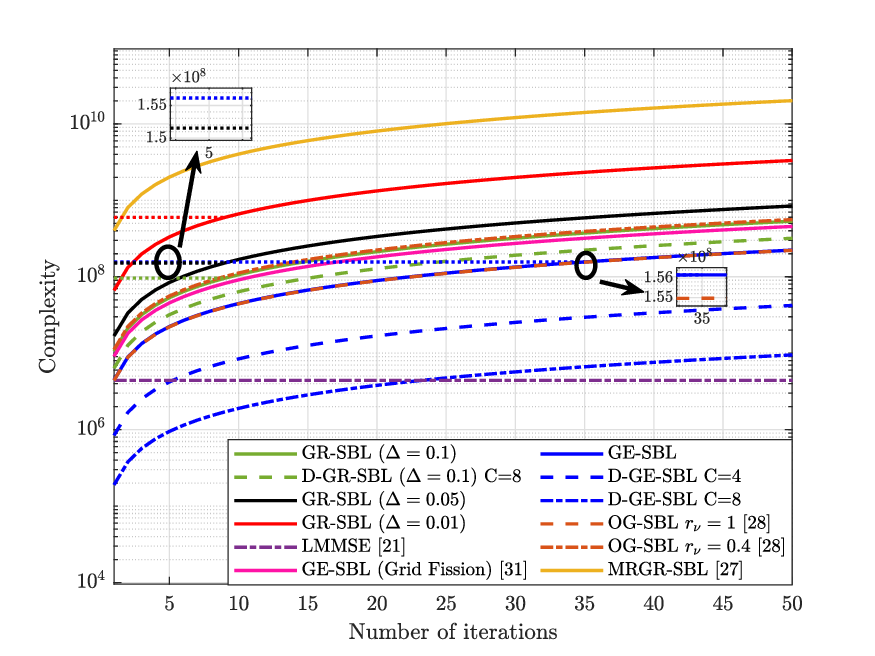}}
	\\
	\subfloat[Average run time comparison of different estimators.]{\includegraphics[scale=0.55]{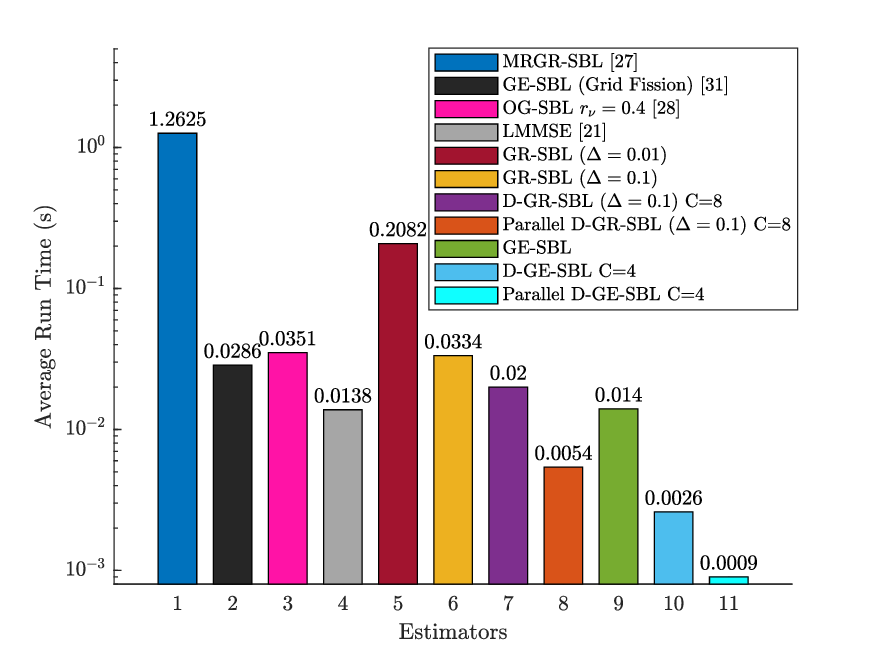}}
	\hfil
	\caption{Complexity and average run time comparison of different estimators.}
	\label{Fig_7}
\end{figure}

We then provide the complexity analysis of different estimators in Fig. \ref{Fig_7}(a). It can be observed that the LMMSE \cite{b18} shows an appropriate complexity level while the complexity of OG-SBL \cite{b25} experiences a significant increase with improved
Doppler sampling resolution $r_{\nu}$. The GE-SBL based on grid fission \cite{b27.1} exhibits an increased complexity compared to OG-SBL with $r_{\nu} = 1$ due to the expanded measurement matrix dimension from grid fission. 
The MRGR-SBL \cite{b24} with a local interpolated high-resolution grid presents a prohibitive complexity due to significant expanded measurement matrix dimension.  
Then, the complexity significantly increases with a finer search step size for the proposed GR-SBL estimator. Although the GR-SBL with fine step size $\Delta = 0.01$ shows a stronger performance advantage, it exhibits a significantly computational burden. 
The GE-SBL estimator shows a lower complexity than GR-SBL with the same number of iterations.
To further conduct a fair comparison, we evaluate their complexity based on the number of iterations required to achieve performance convergence, which are depicted with the horizontal dotted lines in Fig. \ref{Fig_7}(a).
For example, the number of iterations required to achieve the performance convergence for the GR-SBL estimator is approximately 9 at SNR = 20dB, and the GE-SBL estimator  requires approximately 35 iterations observed from Fig. \ref{Fig_6}. We can observe that 
the GE-SBL shows a lower complexity than GR-SBL with step size $\Delta = 0.01$ and better performance than GR-SBL with step size $\Delta = 0.1$, exhibiting a good trade-off between performance and complexity. 
The proposed distributed computing schemes also show an obvious complexity advantage, where the complexity further decreases as $C$ increases. This is due to the fact that smaller dimensional sub-models are processed by the SBL framework, leading to a smaller dimensional matrix inverse. 

To evaluate the practicality of different estimators, we further analyze the average run time\footnote{\label{n3} Note that the average run time of all algorithms are measured on Intel(R) Core(TM) i7-10510U CPU @ 1.80GHz processor. Specifically, we perform parallel processing based on four physical core and other algorithms based on serial modes using single physical core.} in Fig. \ref{Fig_7}(b). 
It can be observed that the average run time results based on serial modes in Fig. \ref{Fig_7}(b) align well with the complexity analysis in Fig. \ref{Fig_7}(a), demonstrating the effectiveness of our analysis regarding computational complexity. Furthermore, the average run time of the proposed D-GR-SBL and D-GE-SBL based on parallel computing show a significantly reduced processing latency, which enables the practical implementation of the proposed off-grid SBL estimators in AFDM systems.

%\begin{figure}[t]
%	\centering
%	\includegraphics[scale=0.55]{Complexity.eps}
%	\caption{Complexity comparison of different estimators.}
%	\label{Fig_7}
%\end{figure} 
%
%\begin{figure}[t]
%	\centering
%	\includegraphics[scale=0.55]{Run_time.eps}
%	\caption{Complexity comparison of different estimators.}
%	\label{Fig_7.1}
%\end{figure}

\begin{figure}[t]
	\centering
	\includegraphics[scale=0.55]{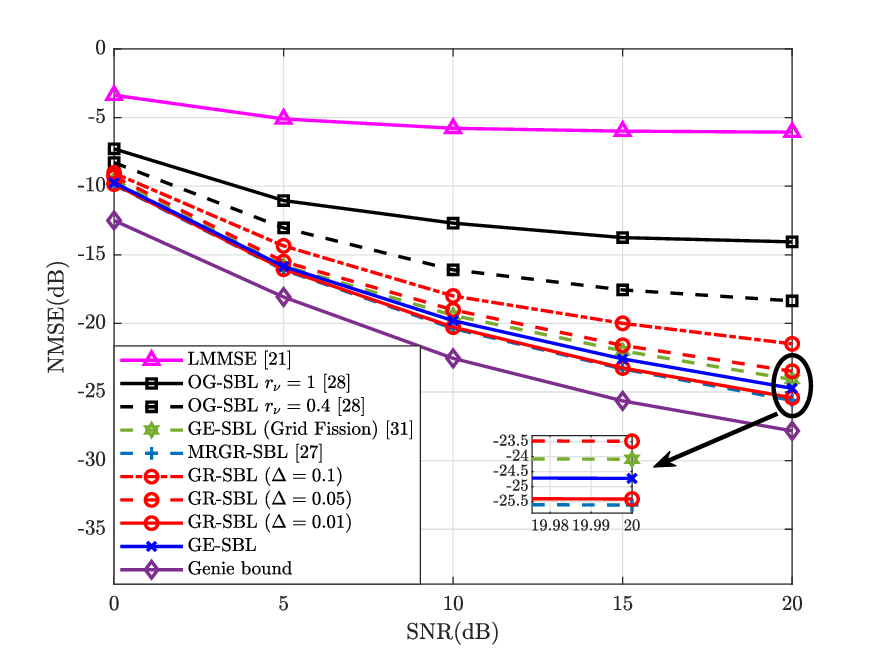}
	\caption{NMSE performance analysis for different channel estimators.}
	\label{Fig_8}
\end{figure}

We now compare the NMSE performance between different channel estimators in Fig. \ref{Fig_8}. The LMMSE \cite{b18} first shows the worst performance. Although the OG-SBL \cite{b25} shows better estimation accuracy than LMMSE by extracting off-grid component based on the first-order linear approximation, the linear approximation errors based on the fixed grid lead to a significant performance degradation. The MRGR-SBL \cite{b24} shows an impressive performance and the GE-SBL based on grid fission \cite{b27.1} also shows a considerable performance, but both of them suffer the increased complexity due to expanded measurement matrix dimension observed from Fig. \ref{Fig_7}.
To overcome these challenges, our proposed GR-SBL and GE-SBL estimators employ a dynamic grid update strategy for estimation performance improvement. Specifically, the GR-SBL shows a significant performance advantage with finer step size. 
However, the finer step size will result in a higher computational burden observed from Fig. \ref{Fig_7}.
For this issue, the proposed GE-SBL estimator overcomes the dependence on the search step size by
performing grid evolution to ensure that the off-grid channel taps can finally match the updated virtual grid, achieving a better trade-off between the performance and complexity. It can be observed that the genie bound achieves the best performance due to the perfect dynamic grid update. Furthermore, the estimation performance of our proposed GR-SBL with a fine step size and that of the GE-SBL are closer to the genie bound, which demonstrates the robustness of our proposed estimators.

\begin{figure}[t]
	\centering
	\includegraphics[scale=0.55]{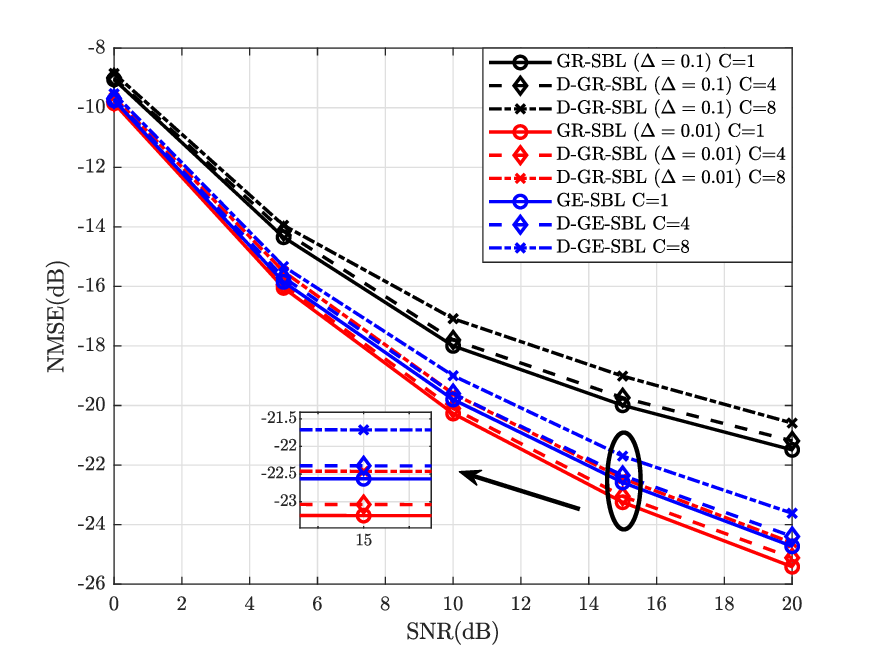}
	\caption{NMSE performance for distributed channel estimators.}
	\label{Fig_9}
\end{figure}

Fig. \ref{Fig_9} shows the NMSE performance for the proposed distributed computing schemes. 
It can be observed that the performance of the proposed 
D-GR-SBL and D-GE-SBL estimators can approach to those of  GR-SBL and GE-SBL estimators with smaller $C$. With the increase of $C$, there is only a slight performance loss but with significant complexity reduction observed from Fig. \ref{Fig_7}. Therefore, our proposed D-GR-SBL and D-GE-SBL estimators can effectively reduce the complexity of off-grid SBL framework and maintain comparable performance with GR-SBL and GE-SBL estimators.

\begin{figure}[t]
	\centering
	\includegraphics[scale=0.55]{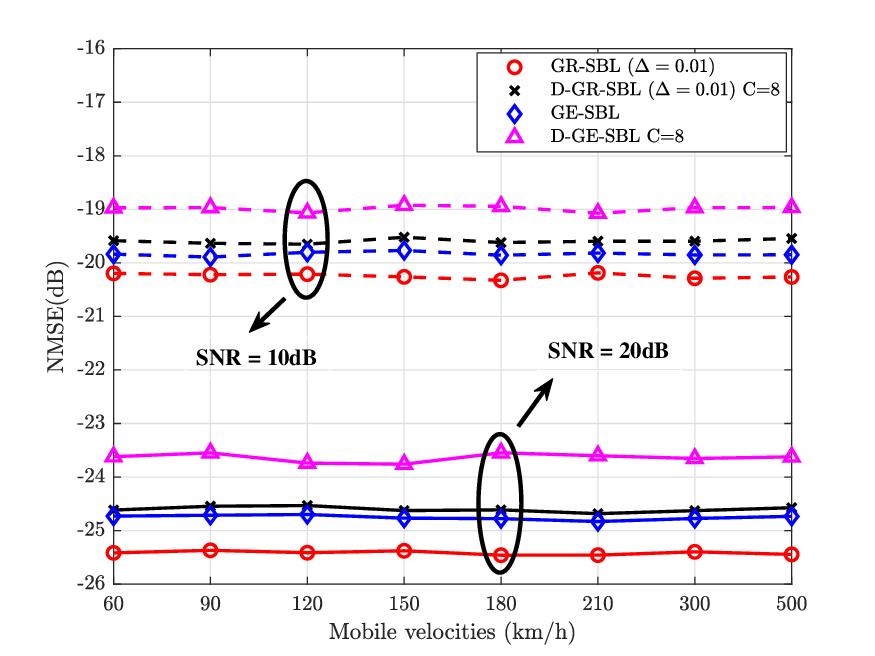}
	\caption{NMSE performance analysis at different mobile velocities.}
	\label{Fig_9.1}
\end{figure}

Fig. \ref{Fig_9.1} shows the NMSE performance of the proposed estimators at different mobile velocities with SNR = 10dB and 20dB. It can be observed that the proposed GR-SBL, GE-SBL and the corresponding distributed computing schemes consistently maintain performance advantages as mobile velocities increase, which strongly supports the robustness of the proposed estimators in high-mobility scenarios.

%\textcolor{blue}{We further test the NMSE performance of the proposed estimators with different mobile velocities in Fig. \ref{Fig_9.1}. It can be observed that the proposed GR-SBL, GE-SBL and the corresponding distributed computing schemes consistently maintain superior performance advantages at different mobile velocities, which strongly supports the robustness of the proposed estimators in high-mobility scenarios.}

\begin{figure}[t]
	\centering
	\includegraphics[scale=0.55]{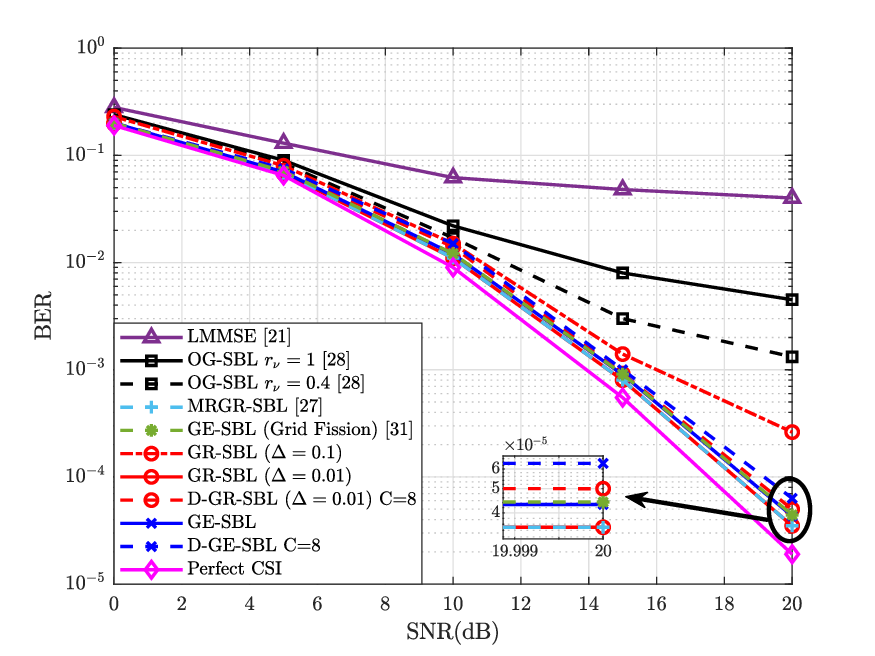}
	\caption{BER performance analysis for different channel estimators.}
	\label{Fig_10}
\end{figure}

Finally, we evaluate the bit error rate (BER) performance with the orthogonal approximate message passing (OAMP) \cite{b34} detector to demonstrate the effectiveness of the proposed channel estimators in Fig. \ref{Fig_10}. It can be observed that the BER performance based on different channel estimators exhibit a similar trend to that of the channel estimation results, i.e., the more accurate channel estimation provides the better BER performance. Among these estimators, the LMMSE \cite{b18} shows the worst BER performance and OG-SBL \cite{b25} also suffers significant BER degradation due to inaccurate channel estimation. The MRGR-SBL \cite{b24} shows an impressive performance and the GE-SBL based on grid fission \cite{b27.1} also shows a considerable performance, but both of them suffer the increased complexity due to expanded measurement matrix dimension observed from Fig. \ref{Fig_7}.
In contrast, the proposed GR-SBL with a fine step size achieves a comparable performance to MRGR-SBL but lower complexity. The proposed GE-SBL also achieves a similar performance level to GE-SBL based on grid fission while
maintaining lower complexity. Furthermore, the BER performance based on our proposed GR-SBL with fine step size $\Delta = 0.01$ and GE-SBL estimators 
are close to that achieved with perfect CSI.
In addition, it can be observed that 
there is a comparable BER performance for proposed low-complexity distributed estimators, which demonstrates the robustness of our proposed estimators.

\section{Conclusion}
In this paper, we developed a SBL framework for AFDM systems,  and proposed the GR-SBL and GE-SBL estimators to address the off-grid channel estimation challenges over doubly-dispersive channels. The proposed GR-SBL estimator achieved superior estimation accuracy by performing a localized grid refinement to mitigate off-grid estimation error at the cost of an increased search burden. To overcome the search limitation, the proposed GE-SBL estimator extracted the off-grid component based on first-order linear approximation and gradually performed grid evolution to improve estimation performance.
Furthermore, the distributed computing scheme extended GR-SBL and GE-SBL estimators into D-GR-SBL and D-GE-SBL to reduce complexity, while enabling parallel processing to further reduce computational latency with a comparable estimation accuracy. Simulation results demonstrated that the proposed GR-SBL and GE-SBL estimators outperform conventional methods, and the corresponding distributed estimators can effectively reduce the complexity and latency. 
It is worth mentioning that the rapidly changing Doppler spreads, imperfect pilot designs, hardware constraints, the advanced synchronization methods and the impact of imperfect synchronization on channel estimation for AFDM system are interesting future works under practical vehicular scenarios.

\begin{figure*}[hb]
	\hrulefill
	\begin{align}\nonumber
		& {\mathcal{L}} \left( \boldsymbol{\alpha}^{(t)}, \boldsymbol{\bar{S}}^{(t)}\right) = \ln |\boldsymbol{C}|+\operatorname{tr}\left\{\boldsymbol{C}^{-1} {\boldsymbol{R}}_y\right\} + \text{Const} \\ \nonumber
		& = \ln \left| \boldsymbol{C}_{-p}  \right| + \ln \left| 1 + \alpha_p^{(t)} \boldsymbol{\phi}^{\text{H}}(\bar{S}_p^{(t)}) \boldsymbol{C}_{-p}^{-1} \boldsymbol{\phi} (\bar{S}_p^{(t)}) \right|  + \operatorname{tr} \left\{\boldsymbol{C}^{-1}_{-p}  {\boldsymbol{R}}_y\right\} - \operatorname{tr} \left\{\frac{\boldsymbol{C}^{-1}_{-p} \boldsymbol{\phi}(\bar{S}_p^{(t)}) \boldsymbol{\phi}^{\text{H}}(\bar{S}_p^{(t)}) \boldsymbol{C}^{-1}_{-p} {\boldsymbol{R}}_y}{ 1 / \alpha_p^{(t)} + \boldsymbol{\phi}^{\text{H}}(\bar{S}_p^{(t)}) \boldsymbol{C}^{-1}_{-p} \boldsymbol{\phi}(\bar{S}_p^{(t)})} \right\} + \text{Const} \\ \tag{61}
		& = \underbrace{\ln \left( 1 + \alpha_p^{(t)} \boldsymbol{\phi}^{\text{H}}(\bar{S}_p^{(t)}) \boldsymbol{C}_{-p}^{-1} \boldsymbol{\phi} (\bar{S}_p^{(t)}) \right) - \frac{\boldsymbol{\phi}^{\text{H}}(\bar{S}_p^{(t)}) \boldsymbol{C}^{-1}_{-p} {\boldsymbol{R}}_y \boldsymbol{C}^{-1}_{-p} \boldsymbol{\phi}(\bar{S}_p^{(t)})}{1 /\alpha_p^{(t)} + \boldsymbol{\phi}^{\text{H}}(\bar{S}_p^{(t)}) \boldsymbol{C}^{-1}_{-p} \boldsymbol{\phi}(\bar{S}_p^{(t)})}}_{\mathcal{L} \left( \alpha_p^{(t)}, \bar{S}_p^{(t)} \right)}  + \underbrace{\ln \left|\boldsymbol{C}_{-p}\right| + \operatorname{tr}\left\{\boldsymbol{C}_{-p}^{-1} {\boldsymbol{R}}_y\right\}}_{\mathcal{L} \left( \boldsymbol{\alpha}_{-p}^{(t)}, \boldsymbol{\bar{S}}_{-p}^{(t)} \right)} + \text{Const}.
	\end{align}
\end{figure*} 

\begin{figure*}[hb]
	\hrulefill
	\begin{align}\nonumber
		& \mathbb{E}_{\Pr (\boldsymbol{\bar{h}} |\boldsymbol{y}_{\text{T}}; \boldsymbol{\alpha}^{(t)}, \gamma^{(t)}, \boldsymbol{\bar{S}}^{(t)} )} \left\lbrace \left| \left| \boldsymbol{y}_{\text{T}} - \boldsymbol{\tilde{\Phi}} (\boldsymbol{\bar{S}}^{(t)}) \boldsymbol{\bar{h}} \right|\right|^2  \right\rbrace  =  \boldsymbol{\bar{\beta}}^{\text{T}} \underbrace{\mathcal{R} \left\lbrace \left( \boldsymbol{\Psi}^{\text{H}} (\boldsymbol{\bar{S}}^{(t)}) \boldsymbol{\Psi} (\boldsymbol{\bar{S}}^{(t)})\right)^{*}  \odot \left( \boldsymbol{\mu}^{(t)}  \left( \boldsymbol{\mu}^{(t)} \right)^{\text{H}} + \boldsymbol{\Sigma}^{(t)} \right)  \right\rbrace}_{\boldsymbol{A}^{(t)}} \boldsymbol{\bar{\beta}} \\ \tag{65}
		& ~~~~~~~~~~~~~~~~- 2 \underbrace{ \mathcal{R} \left\lbrace \text{diag} \left\lbrace \boldsymbol{\mu}^{(t)}\right\rbrace  \boldsymbol{\Psi}^{\text{H}} (\boldsymbol{\bar{S}}^{(t)}) \left( \boldsymbol{y}_{\text{T}} - \boldsymbol{\Phi} (\boldsymbol{\bar{S}}^{(t)}) \boldsymbol{\mu}^{(t)} \right)  - \text{diag} \left\lbrace \boldsymbol{\Psi}^{\text{H}} (\boldsymbol{\bar{S}}^{(t)}) \boldsymbol{\Phi}(\boldsymbol{\bar{S}}^{(t)}) \boldsymbol{\Sigma}^{(t)} \right\rbrace \right\rbrace^{\text{T}}}_{\left( \boldsymbol{b}^{(t)} \right)^{\text{T}} }
		 \boldsymbol{\bar{\beta}} + \text{Const}.
	\end{align}
\end{figure*}

\renewcommand{\appendixname}{Appendix A}
\appendix
From \eqref{29}, the logarithmic objective function is given by ${\mathcal{L}} \left( \boldsymbol{\alpha}^{(t)}, \boldsymbol{\bar{S}}^{(t)}\right)  = \ln |\boldsymbol{C}|+\operatorname{tr}\left\{\boldsymbol{C}^{-1} {\boldsymbol{R}}_y\right\} + \text{Const}$.
To decouple the influence of $p$-th grid component from ${\mathcal{L}} \left( \boldsymbol{\alpha}^{(t)}, \boldsymbol{\bar{S}}^{(t)}\right)$, we first rewritten matrix $\boldsymbol{C}$ as
\begin{align}\nonumber
	& \boldsymbol{C} = {\boldsymbol{\Phi}} (\boldsymbol{\bar{S}}^{(t)}) \boldsymbol{\Lambda}^{(t)} {\boldsymbol{\Phi}}^{\mathrm{H}} (\boldsymbol{\bar{S}}^{(t)}) + \left( \gamma^{(t)} \right)^{-1}  \boldsymbol{I}_{M_{\text{T}}}\label{55} \\
	& \!=\! \alpha_p^{(t)} \boldsymbol{\phi} (\!\bar{S}_p^{(t)}\!) \boldsymbol{\phi}^{\text{H}} (\!\bar{S}_p^{(t)}\!) \!+\! \underbrace{  {\boldsymbol{\Phi}_{\!-p}} (\!\boldsymbol{\bar{S}}_{\!-p}^{(t)}\!) \boldsymbol{\Lambda}_{\!-p}^{(t)} {\boldsymbol{\Phi}_{\!-p}^{\text{H}}} (\!\boldsymbol{\bar{S}}_{\!-p}^{(t)}\!) \!+\! \left(\! \gamma^{(t)}\!\right)^{\!\!-1} \!\!\! \boldsymbol{I}_{M_{\text{T}}} }_{\boldsymbol{C}_{-p}}.
\end{align}

According to the matrix determinant lemma (Lemma 1.1 in \cite{b35}) and Sherman-Morrison-Woodbury lemma \cite{b36}, $|\boldsymbol{C}|$ and $\boldsymbol{C}^{-1}$ can be further expressed as 
\begin{align}\nonumber
	\left| \boldsymbol{C}\right|  & = \left|  \boldsymbol{C}_{-p} + \alpha_p^{(t)} \boldsymbol{\phi} (\bar{S}_p^{(t)}) \boldsymbol{\phi}^{\text{H}} (\bar{S}_p^{(t)}) \right| \\ \tag{59}
	& = \left| \boldsymbol{C}_{-p}\right| \left| 1 + \alpha_p^{(t)} \boldsymbol{\phi}^{\text{H}}(\bar{S}_p^{(t)}) \boldsymbol{C}_{-p}^{-1} \boldsymbol{\phi} (\bar{S}_p^{(t)}) \right|,
\end{align}
\begin{align}\nonumber
	\boldsymbol{C}^{-1} & = \left( \boldsymbol{C}_{-p} + \alpha_p^{(t)} \boldsymbol{\phi} (\bar{S}_p^{(t)}) \boldsymbol{\phi}^{\text{H}} (\bar{S}_p^{(t)}) \right)^{-1} \\ \tag{60}
	& = \boldsymbol{C}_{-p}^{-1} - \frac{\boldsymbol{C}_{-p}^{-1} \boldsymbol{\phi} (\bar{S}_p^{(t)})  \boldsymbol{\phi}^{\text{H}}(\bar{S}_p^{(t)}) \boldsymbol{C}_{-p}^{-1}  }{ 1 / \alpha_p^{(t)} + \boldsymbol{\phi}^{\text{H}}(\bar{S}_p^{(t)}) \boldsymbol{C}_{-p}^{-1} \boldsymbol{\phi} (\bar{S}_p^{(t)})}.
\end{align}

Based on (59) and (60), the logarithmic objective function ${\mathcal{L}} \left( \boldsymbol{\alpha}^{(t)}, \boldsymbol{\bar{S}}^{(t)}\right)$ can be further expressed as (61), shown at the bottom of next page, which mainly consists of the influence of $p$-th grid component $\mathcal{L} \left( \alpha_p^{(t)}, \bar{S}_p^{(t)} \right)$ and the influence of other grid components without $p$-th gird component $\mathcal{L} \left( \boldsymbol{\alpha}_{-p}^{(t)}, \boldsymbol{\bar{S}}_{-p}^{(t)} \right) $.

\renewcommand{\appendixname}{Appendix B}
\appendix
From \eqref{41}, the update of $\boldsymbol{\bar{\beta}}^{(t+1)}$ primarily depends on $\mathbb{E}_{\Pr (\boldsymbol{\bar{h}} |\boldsymbol{y}_{\text{T}}; \boldsymbol{\alpha}^{(t)}, \gamma^{(t)}, \boldsymbol{\bar{S}}^{(t)} )} \left\lbrace \left| \left| \boldsymbol{y}_{\text{T}} - \boldsymbol{\tilde{\Phi}} (\boldsymbol{\bar{S}}^{(t)}) \boldsymbol{\bar{h}} \right|\right|^2  \right\rbrace$, which can be further expressed as
\begin{align}\nonumber
	 & \mathbb{E}_{\Pr (\boldsymbol{\bar{h}} |\boldsymbol{y}_{\text{T}}; \boldsymbol{\alpha}^{(t)}, \gamma^{(t)}, \boldsymbol{\bar{S}}^{(t)} )} \left\lbrace \left| \left| \boldsymbol{y}_{\text{T}} - \boldsymbol{\tilde{\Phi}} (\boldsymbol{\bar{S}}^{(t)}) \boldsymbol{\bar{h}} \right|\right|^2  \right\rbrace \\ \tag{62} 
	& = \left| \left| \boldsymbol{y}_{\text{T}} \!-\! \boldsymbol{\tilde{\Phi}} (\boldsymbol{\bar{S}}^{(t)}) \boldsymbol{\mu}^{(t)} \right|\right|^2 \!\!+\! \operatorname{tr} \!\left\lbrace\! \boldsymbol{\tilde{\Phi}} (\boldsymbol{\bar{S}}^{(t)}) \boldsymbol{\Sigma}^{(t)} \boldsymbol{\tilde{\Phi}}^{\text{H}} (\boldsymbol{\bar{S}}^{(t)}) \!\right\rbrace\!.
\end{align}

By substituting \eqref{39} into (62), the first term of (62) $\left| \left| \boldsymbol{y}_{\text{T}} - \boldsymbol{\tilde{\Phi}} (\boldsymbol{\bar{S}}^{(t)}) \boldsymbol{\mu}^{(t)} \right|\right|^2$ can be further expressed as 
\begin{align}\nonumber
	& \left| \left| \boldsymbol{y}_{\text{T}} - \boldsymbol{\tilde{\Phi}} (\boldsymbol{\bar{S}}^{(t)}) \boldsymbol{\mu}^{(t)} \right|\right|^2 \\ \nonumber
	& = \left| \left| \boldsymbol{y}_{\text{T}} - \left( \boldsymbol{\Phi} (\boldsymbol{\bar{S}}^{(t)})  + \boldsymbol{\Psi} (\boldsymbol{\bar{S}}^{(t)})   \text{diag} \{\boldsymbol{\bar{\beta}}\} \right)  \boldsymbol{\mu}^{(t)} \right|\right|^2 \\ \nonumber
	& =  \boldsymbol{\bar{\beta}}^{\text{T}}  \mathcal{R} \left\lbrace \left[ \left( \boldsymbol{\Psi}^{\text{H}} (\boldsymbol{\bar{S}}^{(t)}) \boldsymbol{\Psi} (\boldsymbol{\bar{S}}^{(t)})\right)^{*}  \odot \left(  \boldsymbol{\mu}^{(t)}  \left( \boldsymbol{\mu}^{(t)} \right)^{\text{H}}  \right)  \right]  \right\rbrace  \boldsymbol{\bar{\beta}} \\ \tag{63}
	& -\! 2 \mathcal{R} \! \left\lbrace \!\text{diag} \{(\boldsymbol{\mu}^{(t)})^{*\!}\} \boldsymbol{\Psi}^{\text{H}} (\boldsymbol{\bar{S}}^{(t)})  \!\left(\! \boldsymbol{y}_{\text{T}} \!-\! \boldsymbol{\Phi} (\boldsymbol{\bar{S}}^{(t)})\boldsymbol{\mu}^{(t)} \! \right) \!\right\rbrace^{\text{T}} \!\!  \boldsymbol{\bar{\beta}} \!+\! \text{Const}.
\end{align}

The second term of (62) $\operatorname{tr} \left\lbrace \boldsymbol{\tilde{\Phi}} (\boldsymbol{\bar{S}}^{(t)}) \boldsymbol{\Sigma}^{(t)} \boldsymbol{\tilde{\Phi}}^{\text{H}} (\boldsymbol{\bar{S}}^{(t)})  \right\rbrace$ can be further expressed as 
\begin{align}\nonumber
	& \operatorname{tr} \left\lbrace \boldsymbol{\tilde{\Phi}} (\boldsymbol{\bar{S}}^{(t)}) \boldsymbol{\Sigma}^{(t)} \boldsymbol{\tilde{\Phi}}^{\text{H}} (\boldsymbol{\bar{S}}^{(t)})  \right\rbrace \\ \nonumber
	& = \operatorname{tr} \left\lbrace \left( \boldsymbol{\Phi} (\boldsymbol{\bar{S}}^{(t)})  + \boldsymbol{\Psi} (\boldsymbol{\bar{S}}^{(t)})   \text{diag} \{\boldsymbol{\bar{\beta}}\} \right)  \boldsymbol{\Sigma}^{(t)} \right. \\ \nonumber
	& ~~~\times \left. \left(  \boldsymbol{\Phi} (\boldsymbol{\bar{S}}^{(t)})  + \boldsymbol{\Psi} (\boldsymbol{\bar{S}}^{(t)})   \text{diag} \{\boldsymbol{\bar{\beta}}\} \right)^{\text{H}}   \right\rbrace \\ \nonumber
	& = \operatorname{tr} \left\lbrace \text{diag} \{\boldsymbol{\bar{\beta}}\} \boldsymbol{\Sigma}^{(t)} \text{diag} \{\boldsymbol{\bar{\beta}}\} \boldsymbol{\Psi}^{\text{H}} (\boldsymbol{\bar{S}}^{(t)}) \boldsymbol{\Psi} (\boldsymbol{\bar{S}}^{(t)}) \right\rbrace \\ \nonumber
	& ~~~+ 2 \mathcal{R} \left\lbrace \operatorname{tr}  \left\lbrace \boldsymbol{\Psi}^{\text{H}} (\boldsymbol{\bar{S}}^{(t)}) \boldsymbol{\Phi} (\boldsymbol{\bar{S}}^{(t)}) \boldsymbol{\Sigma}^{(t)} \text{diag} \{\boldsymbol{\bar{\beta}}\} \right\rbrace  \right\rbrace  + \text{Const} \\ \nonumber
	& = \boldsymbol{\bar{\beta}}^{\text{T}} \mathcal{R} \left\lbrace \boldsymbol{\Sigma}^{(t)} \odot \left( \boldsymbol{\Psi}^{\text{H}} (\boldsymbol{\bar{S}}^{(t)}) \boldsymbol{\Psi} (\boldsymbol{\bar{S}}^{(t)})\right)^{*} \right\rbrace \boldsymbol{\bar{\beta}} \\ \tag{64}
	& ~~~+ 2 \mathcal{R} \left\lbrace \text{diag} \left\lbrace \boldsymbol{\Psi}^{\text{H}} (\boldsymbol{\bar{S}}^{(t)}) \boldsymbol{\Phi} (\boldsymbol{\bar{S}}^{(t)}) \boldsymbol{\Sigma}^{(t)} \right\rbrace  \right\rbrace^{\text{T}} \boldsymbol{\bar{\beta}} + \text{Const}.
\end{align}

By substituting (63) and (64) into (62), $\mathbb{E}_{\Pr (\boldsymbol{\bar{h}} |\boldsymbol{y}_{\text{T}}; \boldsymbol{\alpha}^{(t)}, \gamma^{(t)}, \boldsymbol{\bar{S}}^{(t)} )} \left\lbrace \left| \left| \boldsymbol{y}_{\text{T}} - \boldsymbol{\tilde{\Phi}} (\boldsymbol{\bar{S}}^{(t)}) \boldsymbol{\bar{h}} \right|\right|^2  \right\rbrace$ can be finally expressed as (65), show at the bottom of this page.

%\begin{align}\label{25}
%	& {\mathcal{L}} \left( \boldsymbol{\alpha}^{(t)}, \boldsymbol{\bar{S}}^{(t)}\right) = \ln |\boldsymbol{C}|+\operatorname{tr}\left\{\boldsymbol{C}^{-1} {\boldsymbol{R}}_y\right\} + \text{Const}, \\
%	& = \ln \left| \boldsymbol{C}_{-p}\right| + \ln \left| 1 + \alpha_p^{(t)} \boldsymbol{\phi}^{\text{H}}(\bar{S}_p^{(t)}) \boldsymbol{C}_{-p}^{-1} \boldsymbol{\phi} (\bar{S}_p^{(t)}) \right| \\
%	& + \operatorname{tr} \left\{\boldsymbol{C}^{-1}_{-p}  \right\} + \operatorname{tr} \left\{\frac{\boldsymbol{C}^{-1}_{-p} \boldsymbol{\phi}(\bar{S}_p^{(t)}) \boldsymbol{\phi}^{\text{H}}(\bar{S}_p^{(t)}) \boldsymbol{C}^{-1}_{-p} {\boldsymbol{R}}_y}{ \left( \alpha_p^{(t)}\right)^{-1} + \boldsymbol{\phi}^{\text{H}}(\bar{S}_p^{(t)}) \boldsymbol{C}^{-1}_{-p} \boldsymbol{\phi}(\bar{S}_p^{(t)})} \right\} \\
%	& =  \ln \left( 1 + \alpha_p^{(t)} \boldsymbol{\phi}^{\text{H}}(\bar{S}_p^{(t)}) \boldsymbol{C}_{-p}^{-1} \boldsymbol{\phi} (\bar{S}_p^{(t)}) \right) - \frac{\boldsymbol{\phi}^{\text{H}}(\bar{S}_p^{(t)}) \boldsymbol{C}^{-1}_{-p} {\boldsymbol{R}}_y \boldsymbol{C}^{-1}_{-p} \boldsymbol{\phi}(\bar{S}_p^{(t)})}{\left( \alpha_p^{(t)}\right)^{-1} + \boldsymbol{\phi}^{\text{H}}(\bar{S}_p^{(t)}) \boldsymbol{C}^{-1}_{-p} \boldsymbol{\phi}(\bar{S}_p^{(t)})} \\
%	& + \underbrace{\ln \left|\boldsymbol{C}_{-p}\right| + \operatorname{tr}\left\{\boldsymbol{C}_{-p}^{-1} {\boldsymbol{R}}_y\right\}}_{	\mathcal{L} \left( \boldsymbol{\alpha}_{-p}^{(t)}, \boldsymbol{\bar{S}}_{-p}^{(t)} \right)}
%\end{align}

\vfill

\end{document}